\documentclass{article}
\usepackage{eusipco}                      % Default: 10pt Times Roman  Fonts
\usepackage{amsmath,amssymb,color}
\usepackage[dvips]{graphicx}
\graphicspath{{Eps/}}

\def\babs{\begin{abstract}}             \def\eabs{\end{abstract}}
\def\barr{\begin{array}}                \def\earr{\end{array}}
\def\bcc{\begin{center}}                \def\ecc{\end{center}}

\def\bdes{\begin{description}}          \def\edes{\end{description}}
\def\bdoc{\begin{document}}             \def\edoc{\end{document}}
\def\ben{\begin{enumerate}}             \def\een{\end{enumerate}}
\def\beqn{\begin{eqnarray}}             \def\eeqn{\end{eqnarray}}
\def\beqnl#1{\beqn\label{#1}}           \def\eeqnl#1{\label{#1}\eeqn}
\def\beqnx{\begin{eqnarray*}}           \def\eeqnx{\end{eqnarray*}}
\def\bseqn{\begin{subeqnarray}}         \def\eseqn{\end{subeqnarray}}
\def\beq#1\eeq{\begin{equation}#1\end{equation}}
\def\bal#1\eal{\begin{align}#1\end{align}}
\def\balx#1\ealx{\begin{align*}#1\end{align*}}
\def\beqx{$$}                           \def\eeqx{$$}
\def\bfig{\protect\begin{figure}}       \def\efig{\protect\end{figure}}
\def\bfigx{\protect\begin{figure*}}     \def\efigx{\protect\end{figure*}}
\def\bfigt{\protect\begin{figurette}}   \def\efigt{\protect\end{figurette}}
\def\bfl{\begin{flushleft}}             \def\efl{\end{flushleft}}
\def\bfr{\begin{flushright}}            \def\efr{\end{flushright}}
\def\bit{\begin{itemize}}               \def\eit{\end{itemize}}
\def\bmi{\begin{minipage}}              \def\emi{\end{minipage}}
\def\bfmi{\begin{fminipage}}            \def\efmi{\end{fminipage}}
\def\bpic{\begin{picture}}              \def\epic{\end{picture}}
\def\bqu{\begin{quote}}                 \def\equ{\end{quote}}
\def\bqun{\begin{quotation}}            \def\equn{\end{quotation}}
\def\bsl{\begin{slide}}                 \def\esl{\end{slide}}
\def\btabb{\begin{tabbing}}             \def\etabb{\end{tabbing}}
\def\btabl{\begin{table}}               \def\etabl{\end{table}}
\def\btablx{\begin{table*}}             \def\etablx{\end{table*}}
\def\btab{\begin{tabular}} %\def\etab{\end{tabular}} Conflit avec eta gras... Elle est pas grasse, ma soeur !
\def\btabu{\begin{tabular}}             \def\etabu{\end{tabular}}
\def\btabx{\begin{tabular*}}            \def\etabx{\end{tabular*}}
\def\bbib{\begin{thebibliography}{999}} \def\ebib{\end{thebibliography}}
\def\bver{\begin{verbatim}}             \def\ever{\end{verbatim}}
\def\bca{\begin{cases}}                  \def\eca{\end{cases}}
   
\input{Inputs/alphabet}   
\input{Inputs/abrege}     
\def\pth#1{\left(#1\right)}                \def\stdpth#1{(#1)}
\def\acc#1{\left\{#1\right\}}              \def\stdacc#1{\{#1\}}
\def\cro#1{\left[#1\right]}                \def\stdcro#1{[#1]}
\def\bars#1{\left|#1\right|}               \def\stdbars#1{|#1|}
\def\norm#1{\left\|#1\right\|}             \def\stdnorm#1{\|#1\|}
\def\scal#1{\left\langle#1\right\rangle}   \def\stdscal#1{\langle#1\rangle}
 
\def\bigpth#1{\bigl(#1\bigr)}              \def\biggpth#1{\biggl(#1\biggr)}
\def\bigacc#1{\bigl\{#1\bigr\}}            \def\biggacc#1{\biggl\{#1\biggr\}}
\def\bigcro#1{\bigl[#1\bigr]}              \def\biggcro#1{\biggl[#1\biggr]}
\def\bigbars#1{\bigl|#1\bigr|}             \def\biggbars#1{\biggl|#1\biggr|}
\def\bignorm#1{\bigl\|#1\bigr\|}           \def\biggnorm#1{\biggl\|#1\biggr\|}
\def\bigscal#1{\bigl\langle#1\bigr\rangle} \def\biggscal#1{\biggl\langle#1\biggr\rangle}

\def\Bigpth#1{\Bigl(#1\Bigr)}              \def\Biggpth#1{\Biggl(#1\Biggr)}
\def\Bigacc#1{\Bigl\{#1\Bigr\}}            \def\Biggacc#1{\Biggl\{#1\Biggr\}}
\def\Bigcro#1{\Bigl[#1\Bigr]}              \def\Biggcro#1{\Biggl[#1\Biggr]}
\def\Bigbars#1{\Bigl|#1\Bigr|}             \def\Biggbars#1{\Biggl|#1\Biggr|}
\def\Bignorm#1{\Bigl\|#1\Bigr\|}           \def\Biggnorm#1{\Biggl\|#1\Biggr\|}
\def\Bigscal#1{\Bigl\langle#1\Bigr\rangle} \def\Biggscal#1{\Biggl\langle#1\Biggr\rangle}

\def\diag{{\mathrm{diag}}}              \def\Diag#1{{\mathrm{diag}}\bigcro{#1}} \def\Diagold#1{{\mathrm{diag}}\cro{#1}}
\def\tr{{\mathrm{tr}}\,}                \def\Tr#1{{\mathrm{tr}}\bigcro{#1}}                     \def\Trold#1{{\mathrm{tr}}\cro{#1}}
\def\rg{{\mathrm{rg}}\,}                \def\Rg#1{{\mathrm{rg}}\bigcro{#1}}                     \def\Rgold#1{{\mathrm{rg}}\cro{#1}}
\def\esp{{\mathrm{E}}\,}              \def\Esp#1{{\mathrm{E}}\bigcro{#1}}  \def\Esph#1#2{\underset{{#1}}{\mathrm{E}}\bigcro{#2}}               \def\Espold#1{{\mathrm{E}}\cro{#1}} 
\def\var{{\mathrm{var}}\,}              \def\Var#1{{\mathrm{var}}\bigcro{#1}}           \def\Varold#1{{\mathrm{var}}\cro{#1}}
\def\cov{{\mathrm{Cov}}\,}              \def\Cov#1{{\mathrm{Cov}}\bigcro{#1}}           \def\Covold#1{{\mathrm{Cov}}\cro{#1}}

\def\Cos#1{\cos\cro{#1}}
\def\Sin#1{\sin\cro{#1}}
\def\Exp#1{\exp\cro{#1}}
\def\Log#1{\log\cro{#1}}
\def\Ln#1{\ln\cro{#1}}                  
\def\Det#1{\det\bigcro{#1}}
\def\Erf#1{\mathrm{erf}\cro{#1}} 
\def\Erfc#1{\mathrm{erfc}\cro{#1}}
\def\Erfcx#1{\mathrm{erfcx}\cro{#1}}

\def\cond{\textrm{Cond}\,}              \def\Cond#1{\cond\bigpth{#1}}

\def\sinc{{\mathrm{sinc}}\,}    \def\Sinc#1{{\mathrm{sinc}}\bigcro{#1}} \def\Sincold#1{{\mathrm{sinc}}\cro{#1}}
\def\rang{{\mathrm{rang}}\,}    \def\Rang#1{\rang\bigcro{#1}}                                   \def\Rangold#1{\rang\cro{#1}}
\def\ker{\textrm{Ker}\,}                \def\Ker#1{\ker\bigcro{#1}}                                     \def\Kerold#1{\ker\cro{#1}}
\def\img{\textrm{Im}\,}                 \def\Img#1{\img\bigcro{#1}}                                     \def\Imgold#1{\img\cro{#1}}
\def\vect{{\mathrm{Vect}}\,}    \def\Vect#1{\vect\bigcro{#1}}                                   \def\Vectold#1{\vect\cro{#1}}
\def\sgn{{\mathrm{sgn}}}                \def\Sgn#1{\sgn\bigcro{#1}}                                     \def\Sgnold#1{\sgn\cro{#1}}

\def\Pr{\mathop{\textrm{Pr}}}

\def\reel{{\mathrm{Re}}}                \def\Reel#1{{\mathrm{Re}}\cro{#1}}
\def\card{{\mathrm{Card}}}              \def\Card#1{{\mathrm{Card}}\cro{#1}}

\def\sh{{\mathrm{sh}}}                % sin hyperbolique en FR
\def\ch{{\mathrm{ch}}}                % cos hyperbolique en FR
\def\th{{\mathrm{th}}}                % th hyperbolique en FR
\def\coth{{\mathrm{coth}}}            % cot hyperbolique en FR
\def\coeffbin#1#2{\pth{\setlength{\arraycolsep}{.1em}\barr{c}#1\\#2\earr}}
\def\Div{{\mathrm{div}}}                                                                                        % divergence
\def\Rotv{\overrightarrow{\mathop{{\mathrm{rot}}}}}             % rotationnel avec fleche
\def\Gradv{\overrightarrow{\mathop{{\mathrm{grad}}}}}           % gradient avec fleche
\def\IF{\text{if\:}}             \def\SI{\text{si\:}}
\def\If{\text{If\:}}             \def\Si{\text{Si\:}}
\def\AND{\text{and\:}}           \def\ET{\text{et\:}}
\def\OR{\text{or\:}}             \def\OU{\text{ou\:}}
\def\THEN{\text{then\:}}         \def\ALORS{\text{alors\:}}
                                 \def\DOU{\text{d'o\`u\:}}
\def\WHERE{\text{where\:}}       \def\Ou{\text{o\`u\:}}
\def\WHEN{\text{when\:}}         \def\QUAND{\text{quand\:}}
\def\FOR{\text{for\:}}           \def\POUR{\text{pour\:}}
\def\FORALL{\text{for all\:}}    \def\POURTOUT{\text{pour tout\:}}
\def\ST{\text{s.t.\:}}           \def\SC{\text{s.c.\:}}
\def\SUBJTO{\text{subject to\:}} \def\SOUSC{\text{sous contraintes\:}}
\def\OTHERWISE{\text{otherwise}} \def\SINON{\text{sinon}}
\def\WITH{\text{with\:}}         \def\AVEC{\text{avec\:}}
\def\IN{\text{in\:}}             \def\DANS{\text{dans\:}}

\def\arrayp{\renewcommand{\arraystretch}{.7}\setlength{\arraycolsep}{2pt}}
\def\tabp{\renewcommand{\arraystretch}{.7}\setlength{\tabcolsep}{2pt}}

\newsavebox{\fminibox}
\newlength{\fminilength}
\newenvironment{fminipage}[1][\linewidth]
  {\setlength{\fminilength}{#1}%-2\fboxsep-2\fboxrule}%
   \begin{lrbox}{\fminibox}\begin{minipage}{\fminilength}}
  {\end{minipage}\end{lrbox}\noindent\fbox{\usebox{\fminibox}}}

\def\M{^{-1}} \def\T{^\tD} \def\+{^\dagger}
\def\I{\,|\,}           % "sachant" bien espac\'e pour les formules
\def\J{\mathop{\,;\,}}  % "point virgule" bien espac\'e pour les formules.
\def\w{,\thinspace}
\def\W{,\thickspace}
\def\ldotsv{,\,\ldots,\,}
\def\V{,}               % Virgule pour les nombres decimaux
\def\e#1{.10^{#1}}      % Notation scientifique a la francaise

\def\egdef{\stackrel{\Delta}{=}}
\def\nequiv{\not\kern-.05em\equiv}
\def\egal{\kern-.5em=\kern-.5em}        % Moins d'espace autour de "="
\def\propt{\kern-.2em\propto\kern-.2em} % Idem
\def\dans{\in\!}                        % Trop d'espace apres "appartient a"
\def\pourtt{\forall\,}                  % Pas assez d'espace
\def\wh#1{\widehat{#1}}                 % Sombrero !
\def\wt#1{\widetilde{#1}} 

\def\argmax{\mathop{\mathrm{arg\,max}}} % Mieux que \def\argmax{\arg\max}
\def\argmin{\mathop{\mathrm{arg\,min}}} % car l'indice est reparti
\def\Argmax#1#2{\displaystyle \argmax_{#1}\left\{{#2}\right\}} % Ajout \HS
\def\Argmin#1#2{\displaystyle \argmin_{#1}\left\{{#2}\right\}}  % 

\def\EL{\mathrm{EL}}
\def\Vect{\mathop{\text{Vect}}}

\def\froc#1#2{{#1/#2}}                  % Frac en toc
\def\fric#1#2{\frac1{#2}#1}
\def\fracds#1#2{\frac{\displaystyle#1}{\displaystyle#2}}
\def\diff#1#2{{\frac{d#1}{d#2}}}
\def\dd{\,d}                            % doit etre en italique en anglais
\def\derpar#1#2{{\frac{\partial #1}{\partial #2}}}
\def\derpor#1#2{{\froc{\partial #1}{\partial #2}}}
\def\parsec#1#2#3{{\frac{\partial^2 #1}{\partial #2\,\partial #3}}}
\def\parsecd#1#2{{\frac{\partial^2 #1}{{\partial #2}^2}}}
\def\porsec#1#2#3{{\froc{\partial^2 #1}{\partial #2\,\partial #3}}}
\def\porsecd#1#2{{\froc{\partial^2 #1}{{\partial #2}^2}}}
\def\pornd#1#2#3{{\froc{\partial^{#3} #1}{{\partial #2}^{#3}}}}
\def\intdouble{\int\kern-0.3em\int}
\def\inttriple{\int\kern-0.3em\int\kern-0.3em\int}
\def\prods{\mathop{\text{\footnotesize$\displaystyle\prod$}}}

\def\rond#1{\overset{\kern-0.33em~_\circ}{#1}}
\def\rondit[#1]#2{\overset{\kern#1~_\circ}{#2}}

\def\incirc#1{\pscirclebox[framesep=1pt]{\scriptsize#1}}
\def\incircp#1{\pscirclebox[framesep=1pt]{\tiny#1}}
\def\outcirc#1{\raisebox{-3pt}{\huge$\times$}\hspace*{-.45cm}\text{\incirc{#1}}}

\def\bdoc{\begin{document}}             \def\edoc{\end{document}}
\def\babs{\begin{abstract}}             \def\eabs{\end{abstract}}
\def\bcc{\begin{center}}                \def\ecc{\end{center}}
\def\ben{\begin{enumerate}}             \def\een{\end{enumerate}}
\def\bit{\begin{itemize}}               \def\eit{\end{itemize}}
\def\bpic{\begin{picture}}              \def\epic{\end{picture}}
\def\beq{\begin{equation}}              \def\eeq{\end{equation}}
\def\barr{\begin{array}}                \def\earr{\end{array}}
\def\btab{\begin{tabular}}              \def\etab{\end{tabular}}
\def\btabu{\begin{tabular}}             \def\etabu{\end{tabular}}
\def\bcc{\begin{center}}                \def\ecc{\end{center}}
\def\beqn{\begin{eqnarray}}             \def\eeqn{\end{eqnarray}}
\def\beqnx{\begin{eqnarray*}}           \def\eeqnx{\end{eqnarray*}}
\def\bfig{\begin{figure}}               \def\efig{\end{figure}}
\def\bver{\begin{verb}}						 \def\ever{\end{verb}}

\def\d#1{\,\mbox{d} #1}
\def\disp#1{\displaystyle{#1}}

\def\blue#1{{\color{blue}#1}}
\def\red#1{{\color{red}#1}}
\def\green#1{{\color{green}#1}}
\def\magenta#1{{\color{magenta}#1}}
\def\cyan#1{{\color{cyan}#1}}
\def\titre#1{\setlength{\shadowsize}{4pt}\centerline{\shadowbox{\blue{\mbox{~~\sc #1~~}}}}}

\def\iii{\int_{-\infty}^{+\infty}}
\def\izi{\int_{0}^{\infty}}
\def\izpi{\int_{0}^{\pi}}
\def\izdpi{\int_{0}^{2\pi}}
\def\intd{\int\kern-.8em\int}
\def\intt{\int\kern-.8em\int\kern-.8em\int}
\def\intg{\int\kern-1.1em\int}

\def\xh{\widehat{x}}
\def\thetah{\widehat{\theta}}
\def\betah{\widehat{\beta}}

\def\xbh{\widehat{\xb}}
\def\thetabh{\widehat{\thetab}}
\def\betabh{\widehat{\betab}}

\def\xbhk{\widehat{\xb}^{k}}
\def\thetahk{\widehat{\theta}^{k}}
\def\betahk{\widehat{\beta}^{k}}

\def\xbhkp{\widehat{\xb}^{k+1}}
\def\thetahkp{\widehat{\theta}^{k+1}}
\def\betahkp{\widehat{\beta}^{k+1}}

\def\thetabhk{\widehat{\thetab}^{k}}
\def\betabhk{\widehat{\betab}^{k}}

\def\thetabhkp{\widehat{\thetab}^{k+1}}
\def\betabhkp{\widehat{\betab}^{k+1}}

\def\thetamin{\theta_{\mbox{\tiny min}}}
\def\thetamax{\theta_{\mbox{\tiny max}}}
\def\betamin{\beta_{\mbox{\tiny min}}}
\def\betamax{\beta_{\mbox{\tiny max}}}

\def\ra{\rightarrow}
\def\la{\leftarrow}
\def\da{\downarrow}
\def\ua{\uparrow}

\def\Ra{\Rightarrow}
\def\La{\Leftarrow}
\def\Da{\Downarrow}
\def\Ua{\Uparrow}

\def\lra{\longrightarrow}
\def\lla{\longleftarrow}
\def\Lra{\Longrightarrow}
\def\Lla{\longleftarrow}

\def\lrarr{\leftrightarrow}
\def\Lrarr{\Leftrightarrow}
\def\udarr{\updownarrow}
\def\Uparr{\Updownarrow}

\def\expf#1{\exp\left[ {#1} \right]}

\def\argmax#1#2{\arg\max_{#1}\left\{ #2 \right\}}
\def\argmin#1#2{\arg\min_{#1}\left\{ #2 \right\}}

\def\Prob#1{\mbox{Pr}\left\{#1\right\}}
\def\var#1{\mbox{Var}\left\{#1\right\}}
\def\cov#1{\mbox{Cov}\left\{#1\right\}}
\def\corr#1{\mbox{Corr}\left\{#1\right\}}
\def\trace#1{\mbox{Tr}\left\{#1\right\}}
\def\rang#1{\mbox{rang}\left\{#1\right\}}
\def\det#1{\mbox{d\'et}\left\{#1\right\}}

\def\gbu{\underline{\gb}}
\def\fbu{\underline{\fb}}
\def\zbu{\underline{\zb}}
\def\epsilonbu{\underline{\epsilonb}}
\def\thetabu{\underline{\thetab}}

\def\sigmae{{\sigma_{\epsilon}}}
\def\Sigmae{{\Sigmab_{\epsilonb}}}
\def\Sigmaf{{\Sigmab_{\fb}}}
\def\Sigmag{{\Sigmab_{\gb}}}

\def\fbuh{\widehat{\fbu}}
\def\zbuh{\widehat{\fbu}}
\def\thetabuh{\widehat{\thetabu}}

\def\fbh{\widehat{\fb}}
\def\Pbh{\widehat{\Pb}}
\def\Sigmabh{\widehat{\Sigmab}}
\def\Abh{\widehat{\Ab}}
\def\Bbh{\widehat{\Bb}}
\def\thetabh{\widehat{\thetab}}
\def\Rbe{\Rb_{\epsilon}}
\def\Rbeh{\widehat{\Rbe}}

\def\Rjk{{R_j}_k}
\def\fbik{{\fb_i}_k}
\def\fbjk{{\fb_j}_k}

\def\mik{{m_i}_k}
\def\sigmaik{{\sigma_i^2}_k}
\def\Sigmaik{{\Sigmab_i}_k}
\def\alphaik{{\alpha_i}_k}
\def\betaik{{\beta_i}_k}

\def\muik{{\mu_i}_k}
\def\vik{{v_i}_k}

\def\mjk{{m_j}_k}
\def\sigmajk{{\sigma_j^2}_k}
\def\Sigmajk{{\Sigmab_j}_k}
\def\alphajk{{\alpha_j}_k}
\def\betajk{{\beta_j}_k}

\def\mujk{{\mu_j}_k}
\def\vjk{{v_j}_k}
\def\njk{{n_j}_k}

\def\alphae{\alpha^{\epsilon}}
\def\betae{\beta^{\epsilon}}
\def\alphaez{\alpha^{\epsilon}_{0}}
\def\betaez{\beta^{\epsilon}_{0}}
\def\alphaei{\alpha^{\epsilon}_{i}}
\def\betaei{\beta^{\epsilon}_{i}}
\def\alphaeiz{\alpha^{\epsilon}_{i0}}
\def\betaeiz{\beta^{\epsilon}_{i0}}

\def\mbz{{{\mb}_z}}
\def\Sigmabz{{{\Sigmab}_z}}
\def\diag#1{\mbox{diag}\left[#1\right]}

\def\sigmah{\widehat{\sigma}}
\def\fh{\widehat{f}}
\def\sigmaz{\sigma_z}

\def\blue#1{{\color{blue}#1}}
\def\red#1{{\color{red}#1}}
\def\green#1{{\color{green}#1}}

\def\sumi{\sum_{i=1}^{M} }
\def\sumj{\sum_{j=1}^{N} }
\def\lambdah{\wh{\lambda}}
\def\lambdabh{\wh{\lambdab}}

\def\det#1{\hbox{det}\left(#1\right)}
\def\defined{\,\shortstack{$\triangle$\\ =}\,}

\def\zmat{\left[\matrix{0}\right]}
\def\det#1{\hbox{det}(#1)}
\def\th{\widehat{t}}
\def\xh{\widehat{x}}
\def\lambdah{\widehat{\lambda}}
\def\muh{\widehat{\mu}}
\def\sigmah{\widehat{\sigma}}
\def\rhoh{\widehat{\rho}}
\def\betah{\widehat{\beta}}
\def\alphah{\widehat{\alpha}}
\def\tbh{\widehat{\tb}}
\def\mubh{\widehat{\mub}}
\def\sigmabh{\widehat{\sigmab}}
\def\rhobh{\widehat{\rhob}}
\def\oneb{\mbox{\bf 1}}

\def\bm#1{\mbox{\boldmath $#1$}}
\def\zerob{{\bf 0}}
\def\oneb{{\bf 1}}
\def\intg{\int\kern-1.1em\int}
\def\expf#1{\exp\left[ {#1} \right]}

\def\gbu{\underline{\gb}}
\def\fbu{\underline{\fb}}
\def\zbu{\underline{\zb}}
\def\epsilonbu{\underline{\epsilonb}}
\def\uthetab{\underline{\thetab}}

\def\sigmae{{\sigma_{\epsilon}}}
\def\Sigmae{{\Sigma_{\epsilonb}}}
\def\Sigmabe{{\Sigmab_{\epsilonb}}}
\def\Sigmaf{{\Sigmab_{\fb}}}
\def\Sigmag{{\Sigmab_{\gb}}}

\def\fbuh{\widehat{\fbu}}
\def\zbuh{\widehat{\fbu}}
\def\uthetabh{\widehat{\uthetab}}

\def\fbh{\widehat{\fb}}
\def\Sigmabh{\widehat{\Sigmab}}
\def\Abh{\widehat{\Ab}}
\def\thetabh{\widehat{\thetab}}

\def\Rjk{{R_j}_k}
\def\fbik{{\fb_i}_k}
\def\fbjk{{\fb_j}_k}

\def\mik{{m_i}_k}
\def\sigmaik{{\sigma_i^2}_k}
\def\Sigmaik{{\Sigmab_i}_k}
\def\alphaik{{\alpha_i}_k}
\def\betaik{{\beta_i}_k}

\def\muik{{\mu_i}_k}
\def\vik{{v_i}_k}

\def\mjk{{m_j}_k}
\def\sigmajk{{\sigma_j^2}_k}
\def\Sigmajk{{\Sigmab_j}_k}
\def\alphajk{{\alpha_j}_k}
\def\betajk{{\beta_j}_k}

\def\mujk{{\mu_j}_k}
\def\vjk{{v_j}_k}
\def\njk{{n_j}_k}

\def\alphae{\alpha^{\epsilon}}
\def\betae{\beta^{\epsilon}}
\def\alphaez{\alpha^{\epsilon}_{0}}
\def\betaez{\beta^{\epsilon}_{0}}
\def\alphaei{\alpha^{\epsilon}_{i}}
\def\betaei{\beta^{\epsilon}_{i}}
\def\alphaeiz{\alpha^{\epsilon}_{i0}}
\def\betaeiz{\beta^{\epsilon}_{i0}}
\def\alphaeiz{\alpha^{\epsilon}_{i0}}
\def\betaeiz{\beta^{\epsilon}_{i0}}
\def\alphaeiz{\alpha^{\epsilon}_{i0}}
\def\betaeiz{\beta^{\epsilon}_{i0}}

\def\mbz{{{\mb}_z}}
\def\Sigmabz{{{\Sigmab}_z}}

\def\vect{\mbox{vect}}
\def\lra{\longrightarrow}

\def\gir{g_i(\rb)}
\def\fjr{f_j(\rb)}
\def\zjr{z_j(\rb)}
\def\epsilonir{\epsilon_i(\rb)}
\def\gbr{\gb(\rb)}
\def\fbr{\fb(\rb)}
\def\zbr{\zb(\rb)}
\def\epsilonbr{\epsilonb(\rb)}
\def\fbhr{\fbh(\rb)}
\def\Sigmabhr{\Sigmabh(\rb)}

\def\mbzr{\mb_{z(\rb)}}
\def\Sigmabzr{\Sigmab_{z(\rb)}}
\def\Sigmagbzr{{\Sigmab_g}_{z(\rb)}}

\def\mjzjr#1{{m_{#1}}_{z_{#1}(\rb)}}
\def\sigmajzjr#1{{\sigma_{#1}}_{z_{#1}(\rb)}}

\def\Beta{B}
\def\Betab{\bm{\Beta}}
\def\Betabe{\bm{\Beta}_{\epsilonb}}

\def\fh{\widehat{f}}
\def\sigmah{\widehat{\sigma}}
\def\sigmaei{\sigma_{\epsilon_i}^2}

\def\thetah{\widehat{\theta}}
\def\sigmah{\widehat{\sigma}}

\def\Ftxm{\widetilde{F}_{X}(x_-)}
\def\Ftxp{\widetilde{F}_{X}(x_+)}

\def\NU{{\cal V}}

\def\tFx{\widetilde{F}_{X}}

\def\Fx{F_{X}(x)}
\def\fx{f_{X}(x)}

\def\Fxv{F_{X}(x)}
\def\fxv{f_{X}(x)}

\def\Fxi{{F}_{X_i}(x_i)}
\def\fxi{{f}_{X_i}(x_i)}

\def\Gnu{G_{\NU}(\nu)}
\def\gnu{g_{\NU}(\nu)}

\def\Fnu{F_{\NU}(\nu)}
\def\fnu{f_{\NU}(\nu)}

\def\Finnu#1{F^{-1}_{\NU}(#1)}

\def\Gnua#1{G_{\NU}(#1)}
\def\Gnub#1{G_{\NU}^{-1}(#1)}
\def\Gnuh{G_{\NU}^{-1}(1/2)}

\def\Ftx{\widetilde{F}_{X}(x)}
\def\ftx{\widetilde{f}_{X}(x)}

\def\Ftxi{\widetilde{F}_{X_i}(x_i)}
\def\ftxi{\widetilde{f}_{X_i}(x_i)}

\def\Ftxv{\widetilde{F}_{X}(x)}
\def\FtX#1{\widetilde{F}_{X}(#1)}
\def\ftxv{\widetilde{f}_{X}(x)}

\def\fxnu{f_{X|\NU}(x|\nu)}
\def\fxNU{f_{X|\NU}(x|\NU)}
\def\FxNU{F_{X|\NU}(x|\NU)}

\def\fxNUtheta{f_{X|\NU,\theta}(x|\NU,\theta)}
\def\FxNUtheta{F_{X|\NU,\theta}(x|\NU,\theta)}

\def\Fxnu{F_{X|\NU}(x|\nu)}
\def\Fxnui#1{F_{X|\NU}(x|\nu_#1)}

\def\FxN#1{F_{X|\NU}(#1)}
\def\FxNU{F_{X|\NU}(x|\NU)}
\def\Fxnun#1{F_{X|\NU}(x|#1)}
\def\Fxinun#1{F_{X_i|\NU}(x_i|#1)}
\def\fxnu{f_{X|\NU}(x|\nu)}
\def\fxNU{f_{X|\NU}(x|\NU)}

\def\Fxvnu{F_{X|\NU}(x|\nu)}
\def\Fxvnun#1{F_{X|\NU}(x|#1)}
\def\FxvNU{F_{X|\NU}(x|\NU)}

\def\FXvnu{F_{X|\NU}}

\def\fxvnu{f_{X|\NU}(x|\nu)}
\def\fxvNU{f_{X|\NU}(x|\NU)}

\def\FXvNU#1{{F}_{X|\NU}(#1|\NU)}
\def\FXvn#1{{F}_{X|\NU}(x|#1)}

\def\Fxtheta{F_{X|\theta}(x|\theta)}
\def\fxtheta{f_{X|\theta}(x|\theta)}

\def\Fxvtheta{F_{X|\theta}(x|\theta)}
\def\fxvtheta{f_{X|\theta}(x|\theta)}

\def\Ftxtheta{\widetilde{F}_{X|\theta}(x|\theta)}
\def\ftxtheta{\widetilde{f}_{X|\theta}(x|\theta)}

\def\Ftxvtheta{\widetilde{F}_{X|\theta}(x|\theta)}
\def\ftxvtheta{\widetilde{f}_{X|\theta}(x|\theta)}

\def\Fxnutheta{F_{X|\NU,\theta}(x|\nu,\theta)}
\def\fxnutheta{f_{X|\NU,\theta}(x|\nu,\theta)}

\def\ftheta{f_{\Theta}(\theta)}
\def\fthetaxnuz{f_{\Theta|X,\nu_0}(\theta|x,\nu_0)}
\def\fthetax{f_{\Theta|X}(\theta|x)}

\def\Fxvnutheta{F_{X|\NU,\theta}(x|\nu,\theta)}
\def\FxvNUtheta{F_{X|\NU,\theta}(x|\NU,\theta)}
\def\fxvnutheta{f_{X|\NU,\theta}(x|\nu,\theta)}
\def\Ftxvnutheta{\widetilde{F}_{X|\theta}(x|\theta)}
\def\ftxvnutheta{\widetilde{f}_{X|\theta}(x|\theta)}

\def\FxvNUm{F_{X|\NU}(x_-|\NU)}
\def\FxvNUp{F_{X|\NU}(x_+|\NU)}

\def\Fxinu{F_{X_i|\NU}(x_i|\nu)}
\def\fxinu{f_{X_i|\NU}(x_i|\nu)}

\def\Ftxi{\widetilde{F}_{X_i}(x_i)}
\def\ftxi{\widetilde{f}_{X_i}(x_i)}

\def\fxitheta{f_{X_i|\theta}(x_i|\theta)}
\def\ftxitheta{\widetilde{f}_{X_i|\theta}(x_i|\theta)}

\def\thetah{\widehat{\theta}}
\def\thetat{\widetilde{\theta}}

\def\lnuxv{l_{\nu|X}(\nu|x)}

\def\Lnuxva#1{L_{\nu|X}(#1 | x)}
\def\lnuxva#1{l_{\nu|X}(#1 | x)}

\def\Prob#1{\mbox{P}\left(#1\right)}

\def\lthetaxvnuz{l(\theta)}

\def\lthetaxvnuz{l_{\theta|x,\nu_0}(\theta | x,\nu_0)}
\def\fxnuztheta{f_{X|\nu_0,\theta}(x | \nu_0,\theta)}
\def\fxinuztheta{f_{X|\nu_0,\theta}(x_i | \nu_0,\theta)}
\def\fxvtheta{f_{X|\theta}(x | \theta)}
\def\lthetaxv{l_{\theta|X}(\theta | x)}
\def\ltthetaxv{\widetilde{l}_{\theta|X}(\theta | x)}

\def\ltheta{l(\theta )}
\def\lttheta{\widetilde{l}(\theta)}
\def\Lnu{L(\nu )}
\def\Lnui#1{L(\nu_#1)}
\def\Linu{L_i(\nu )}
\def\Linu{L_i(\nu )}
\def\LNU{L(\NU )}
\def\Lin#1{L^{-1}(#1)}
\def\L#1{L(#1)}
\def\Li#1{L_i(#1)}
\def\FNU#1{F_\NU(#1)}
\def\FinNU#1{F^{-1}_\NU(#1)}

\def\ftthetax{\tilde{f}_{\Theta|X}(\theta|x)}
\def\ER{R}

\def\Ndd#1#2#3{{\cal N}\left( #1 ;~ #2, #3 \right)}
\def\Cdd#1#2#3{{\cal C}\left( #1 ;~ #2, #3 \right)}
\def\Sdd#1#2#3#4{{\cal S}\left( #1 ;~ #2, #3, #4 \right)}
\def\Edd#1#2{{\cal E}\left( #1 ;~ #2 \right)}
\def\DEdd#1#2{{\cal DE}\left( #1 ;~ #2 \right)}
\def\Gdd#1#2#3{{\cal G}\left( #1 ;~ #2, #3 \right)}
\def\IGdd#1#2#3{{\cal IG}\left( #1 ;~ #2, #3 \right)}

\def\Xwr{X(\omega,\rb)}
\def\xwr{x(\omega,\rb)}

\def\muzw{\mu_{z}(\omega)}
\def\muzwr{\mu_{z(\rb)}(\omega)}
\def\szw{\sigma_{z}^2(\omega)}
\def\szwr{\sigma_{z(\rb)}^2(\omega)}
\def\pzw{p_{z}(\omega)}
\def\pzwr{p_{z(\rb)}(\omega)}

\def\hszw{\widehat{\sigma_{z}^2}(\omega)}
\def\hmuzw{\widehat{\mu}_{z}(\omega)}
\def\hpzw{\widehat{p}_{z}(\omega)}

\def\muzwrl{\mu_{z(\rb)}(\omega-l)}

\def\pxwrcz{p\left(\xwr ~|~ z\right)}
\def\pxwr{p\left(\xwr\right)}

\def\Xbw{\Xb(\omega)}
\def\xbw{\xb(\omega)}
\def\uXb{\underline{\Xb}}
\def\uxb{\underline{\xb}}
\def\pxbw{p(\xbw)}
\def\puxb{p(\uxb)}
\def\pzb{P(\zb)}

\def\Szw{S_z(\omega)}
\def\mubw{\mub(\omega)}

\def\Srw{S_{\rb}(\omega)}
\def\Swr{S_{\omega}(\rb)}
\def\Sbw{\Sb(\omega)}
\def\sbw{\sb(\omega)}
\def\Sbr{\Sb(\rb)}
\def\mubw{\mub(\omega)}
\def\cov#1{\mbox{cov}[#1]}

\def\pdf{\emph{pdf}}
\def\pmf{\emph{pmf}}
\def\dpdx#1#2{\frac{\partial #1}{\partial #2}}

\def\REM#1{}

\def\XXwr{X(\omega,\rb)}
\def\xxwr{x(\omega,\rb)}

\def\Xwr{X_{\omega}(\rb)}
\def\xwr{x_{\omega}(\rb)}
\def\Xrw{X_{\rb}(\omega)}
\def\xrw{x_{\rb}(\omega)}

\def\Ywr{Y_{\omega}(\rb)}
\def\ywr{y_{\omega}(\rb)}
\def\Yrw{Y_{\rb}(\omega)}
\def\yrw{y_{\rb}(\omega)}

\def\Ewr{E_{\omega}(\rb)}
\def\ewr{\epsilon_{\omega}(\rb)}
\def\Erw{E_{\rb}(\omega)}
\def\erw{\epsilon_{\rb}(\omega)}

\def\Xbr{\Xb(\rb)}
\def\xbr{\xb(\rb)}
\def\Xbw{\Xb(\omega)}
\def\xbw{\xb(\omega)}

\def\uXb{\underline{\Xb}}
\def\uxb{\underline{\xb}}
\def\uXbz{\underline{\Xb}_z}
\def\uxbz{\underline{\xb}_z}

\def\uthetab{\underline{\thetab}}
\def\uthetabz{\underline{\thetab}_z}

\def\Ywr{Y_{\omega}(\rb)}
\def\ywr{y_{\omega}(\rb)}
\def\Yrw{Y_{\rb}(\omega)}
\def\yrw{y_{\rb}(\omega)}
\def\Ybr{\Yb(\rb)}
\def\ybr{\yb(\rb)}
\def\Ybw{\Yb(\omega)}
\def\ybw{\yb(\omega)}
\def\uYb{\underline{\Yb}}
\def\uyb{\underline{\yb}}
\def\uYbz{\underline{\Yb}_z}
\def\uybz{\underline{\yb}_z}

\def\Ewr{E_{\omega}(\rb)}
\def\ewr{\epsilon_{\omega}(\rb)}
\def\Erw{E_{\rb}(\omega)}
\def\erw{\epsilon_{\rb}(\omega)}
\def\Ebr{Eb(\rb)}
\def\ebr{\epsilonb(\rb)}
\def\Ebw{Eb(\omega)}
\def\ebw{\epsilonb(\omega)}
\def\uEb{\underline{\Eb}}

\def\Erwp{E_{\rb}(\omega')}
\def\Xrwp{X_{\rb}(\omega')}

\def\xwmr{x_{\omega-1}(\rb)}
\def\muzwm{\mu_{z}(\omega-1)}

\def\rhoz{\rho_z}
\def\sez{\sigma_{\eta_z}^2}

\def\xwrp{x_{\omega}(\rb')}
\def\xrwp{x_{\rb}(\omega')}
\def\xwpr{x_{\omega'}(\rb)}
\def\xwprp{x_{\omega'}(\rb')}
\def\Xwrp{x_{\omega}(\rb')}
\def\Xwrp{X_{\omega}(\rb')}
\def\Xwpr{X_{\omega'}(\rb)}
\def\Xwprp{X_{\omega'}(\rb')}

\def\sigmaed{\sigmae^2}
\def\sigmaew{\sigmae(\omega)}
\def\sigmaedw{\sigmaed(\omega)}

\def\Sigmabe{\Sigmab_{\epsilon}}
\def\sigmabh{\widehat{\Sigmab}}
\def\xbh{\widehat{\xb}}
\def\zbh{\widehat{\zb}}
\def\qbh{\widehat{\qb}}
\def\sbh{\widehat{\sb}}

\def\argmax#1#2{\arg\max_{#1}\left\{#2\right\}}
\def\argmin#1#2{\arg\min_{#1}\left\{#2\right\}}
\def\espx#1#2{\mbox{E}_{#1}\left\{#2\right\}}
\def\esp#1{\mbox{E}\left\{#1\right\}}
\def\d#1{\mbox{~d}#1}

\def\Tr#1{\mbox{Tr}\left[#1\right]}

\def\syslins#1#2#3#4{
\bpic(150,40)
  \put(5,15){\makebox(15,0){#1}}
  \put(20,15){\vector(1,0){10}}
  \put(30,7){\framebox(30,15){#2}}
  \put(60,15){\vector(1,0){10}}
  \put(75,15){\circle{10}}
  \put(75,15){\makebox(0,0){+}}
  \put(80,15){\vector(1,0){10}}
  \put(125,15){\makebox(15,0){#4}}
  \put(75,30){\vector(0,-1){10}}
  \put(68,32){\makebox(30,0){#3}}
\epic
}
\def\syslinsa#1#2#3#4{
\bpic(90,40)
  \put(5,15){\makebox(10,0){#1}}
  \put(15,15){\vector(1,0){10}}
  \put(25,7){\framebox(20,15){#2}}
  \put(45,15){\vector(1,0){10}}
  \put(60,15){\circle{10}}
  \put(60,15){\makebox(0,0){+}}
  \put(65,15){\vector(1,0){10}}
  \put(75,15){\makebox(10,0){#4}}
  \put(60,30){\vector(0,-1){10}}
  \put(60,32){\makebox(10,0){#3}}
\epic
}
\def\inversions#1#2#3{
\begin{picture}(100,40)
  \put(0,15){\makebox(15,0){#1}}
  \put(20,15){\vector(1,0){10}}
  \put(30,7){\framebox(40,15){#2}}
  \put(70,15){\vector(1,0){10}}
  \put(85,15){\makebox(15,0){#3}}
\end{picture}
}

\def\tomoXaz{
\begin{picture}(70,68)
  \put(15,35){\makebox(20,15){$\red{f(x,y)}$}}
  \put(-1,0){\axexy}
  \put(10,10){\objet}
  \put(5,5){\vector(1,1){55}}
  \put(55,55){\makebox(5,10){$r$}}
  \put(35,28){\makebox(5,10){$\phi$}}
  \put(60,8){\makebox(10,15){$\bullet$D}}
  \put(60,0){\makebox(15,10){$\blue{g(r,\phi)}$}}
  \put(10,55){\makebox(10,10){S$\bullet$}}
  \put(62,15){\line(-1,1){45}}
  \put(40,-5){\proj}
\thinlines  \blue{\multiput(10,14)(5,0){9}{\line(0,1){40}}}
\thinlines  \blue{\multiput(8,14)(0,5){9}{\line(1,0){40}}}
\end{picture}
}

\def\tomoXay{
\begin{picture}(70,70)
  \put(15,35){\makebox(20,15){$f(x,y)$}}
  \put(-1,0){\axexy}
  \put(10,10){\objet}
  \put(5,5){\vector(1,1){55}}
  \put(55,55){\makebox(5,10){$r$}}
  \put(35,28){\makebox(5,10){$\phi$}}
  \put(60,8){\makebox(10,15){$\bullet$D}}
  \put(60,0){\makebox(15,10){$g(r,\phi)$}}
  \put(10,55){\makebox(10,10){S$\bullet$}}
  \put(62,15){\line(-1,1){45}}
  \put(40,-5){\proj}
\end{picture}
}

\def\slice{
\begin{picture}(140,70)
  \put(13,35){\makebox(20,15){\red{$f(x,y)$}}}
  \put(35,28){\makebox(5,10){$\phi$}}
  \put(80,10){\makebox(5,10){\blue{$g(r,\phi)$--FT--$G(\Omega,\phi)$}}}
  \put(0,0){\axexy}
  \put(0,0){\axers}
  \put(10,10){\objet}
  \put(35,28){\makebox(5,10){$\phi$}}
  \put(35,-5){\proj}
  \put(80,0){\axewxwy}
  \put(80,0){\axeaw}
  \put(90,30){\makebox(20,15){\red{$F(\omega_{x}, \omega_{y})$}}}
  \put(115,28){\makebox(5,10){$\phi$}}
  \put(96,15){\line(1,1){30}}
\end{picture}
}

\def\cresta{\tt    \setlength{\unitlength}{0.8pt}
\begin{picture}(120,120)
\thicklines   \multiput(0,27)(0,20){5}{\line(1,0){80}}
              \multiput(0,27)(20,0){5}{\line(0,1){80}}
              \put(61,33){\red{$f_N$}}
              \put(0,92){\red{$f_1$}}
              \put(40,72){\red{$f_j$}}
              \put(88,50){\blue{$g_{i}$}}
              \put(38,120){\blue{$H_{ij}$}}
              \put(-5,115){\line(3,-2){90}}
              \put(42,85){\blue{\line(3,-2){20}}}
\end{picture}}

\def\crestb{\tt    \setlength{\unitlength}{0.8pt}
\begin{picture}(120,120)
\thicklines   \multiput(0,27)(0,20){5}{\line(1,0){80}}
              \multiput(0,27)(20,0){5}{\line(0,1){80}}
              \put(61,33){\red{$f_N$}}
              \put(0,92){\red{$f_1$}}
              \put(40,72){\red{$f_j$}}
              \put(0,12){\blue{$g_1$}}
              \put(60,12){\blue{$g_m$}}
              \put(88,33){\blue{$g_{m+1}$}}
              \put(88,75){\blue{$g_i$}}
              \put(88,95){\blue{$g_M$}}
              \put(-5,75){\blue{\line(1,0){90}}}
\end{picture}}

\def\homedir{/home/djafari/}
\def\Picfs{\homedir Tex/Ali/Picfs/}
\def\Epsfs{\homedir Tex/Ali/Eps/}
\def\sfEps{\homedir Matlab/sf2d/Eps/}
\def\decEps{\homedir Matlab/dec1d/Eps/}
\def\imEps{\homedir Matlab/dec2d/Eps/}
\def\matlabdat{\homedir Matlab/dat/}
\def\radonEps{\homedir Matlab/Radon/Eps/}
\def\Physics{\homedir Tex/Ali/Articles/Physics/}
\def\Maxent{\homedir Tex/Ali/Congres/Maxent96/}
\def\RappEps{\homedir Tex/Ali/Rapports/Rapp95/Eps/}
\def\ProbaEps{\homedir Tex/Ali/Proba/Eps/}

\def\octagone{
\bpic(30,30)
    \put(10,0){\line(1,0){10}}
    \put(20,0){\line(1,1){10}}
    \put(30,10){\line(0,1){10}}
    \put(30,20){\line(-1,1){10}}
    \put(20,30){\line(-1,0){10}}
    \put(10,30){\line(-1,-1){10}}
    \put(0,20){\line(0,-1){10}}
    \put(0,10){\line(1,-1){10}}
\epic
}

\def\octa{
\bpic(30,20)
    \put(5,0){\line(1,0){20}}
    \put(25,0){\line(1,1){5}}
    \put(30,5){\line(0,1){10}}
    \put(30,15){\line(-1,1){5}}
    \put(25,20){\line(-1,0){20}}
    \put(5,20){\line(-1,-1){5}}
    \put(0,15){\line(0,-1){10}}
    \put(0,5){\line(1,-1){5}}
\epic
}

\def\octb{
\bpic(30,20)
    \put(0,0){\octa}
    \put(1,0){\octa}
\epic
}

\def\octaT#1{
\bpic(30,20)
    \put(0,0){\octa}
    \put(0,0){\makebox(35,20){#1}}
\epic
}

\def\octbT#1{
\bpic(30,20)
    \put(0,0){\octb}
    \put(0,0){\makebox(35,20){#1}}
\epic
}

\def\hexagone{
\bpic(30,30)
    \put(15,0){\line(3,2){15}}
    \put(30,10){\line(0,1){10}}
    \put(30,20){\line(-3,2){15}}
    \put(15,30){\line(-3,-2){15}}
    \put(0,20){\line(0,-1){10}}
    \put(0,10){\line(3,-2){15}}
\epic
}

\def\hexa{
\bpic(30,20)
    \put(5,0){\line(1,0){20}}
    \put(25,0){\line(1,2){5}}
    \put(30,10){\line(-1,2){5}}
    \put(25,20){\line(-1,0){20}}
    \put(5,20){\line(-1,-2){5}}
    \put(0,10){\line(1,-2){5}}
\epic
}

\def\hexb{
\bpic(30,20)
    \put(0,0){\hexa}
    \put(1,0){\hexa}
\epic
}

\def\hexaT#1{
\bpic(30,20)
    \put(0,0){\hexa}
    \put(0,0){\makebox(35,20){#1}}
\epic
}

\def\hexbT#1{
\bpic(30,20)
    \put(0,0){\hexb}
    \put(0,0){\makebox(35,20){#1}}
\epic
}

\def\surface{
\bpic(30,30)
    \put(20,0){\line(1,1){10}}
    \put(30,10){\line(-1,1){20}}
    \put(10,30){\line(-1,-1){10}}
    \put(0,20){\line(1,-1){20}}
\epic
}

\def\losange{
\bpic(30,30)
    \put(15,0){\line(1,1){15}}
    \put(30,15){\line(-1,1){15}}
    \put(15,30){\line(-1,-1){15}}
    \put(0,15){\line(1,-1){15}}
\epic
}

\def\axexy{
\bpic(65,65)
  \put(0,30){\vector(1,0){65}}
  \put(60,25){\makebox(5,5){$x$}}
  \put(30,0){\vector(0,1){65}}
  \put(30,60){\makebox(5,5){$y$}}
\epic
}

\def\axexyz{
\bpic(65,65)
  \put(30,30){\vector(1,0){35}}
  \put(60,25){\makebox(5,5){$x$}}
  \put(30,30){\vector(0,1){35}}
  \put(30,60){\makebox(5,5){$y$}}
  \put(30,30){\vector(-1,-2){12}}
  \put(20,5){\makebox(5,5){$z$}}
\epic
}

\def\axewxwy{
\bpic(65,65)
  \put(0,30){\vector(1,0){65}}
  \put(60,25){\makebox(5,5){$\omega_x$}}
  \put(30,0){\vector(0,1){65}}
  \put(31,60){\makebox(5,5){$\omega_y$}}
\epic
}

\def\axers{
\bpic(65,65)
  \put(5,5){\vector(1,1){55}}
  \put(55,55){\makebox(5,10){$r$}}
  \put(55,5){\vector(-1,1){55}}
  \put(0,55){\makebox(5,10){$s$}}
\epic
}

\def\axeab{
\bpic(65,65)
  \put(5,5){\vector(1,1){55}}
  \put(55,55){\makebox(5,10){$\beta$}}
  \put(55,5){\vector(-1,1){55}}
  \put(0,55){\makebox(5,10){$\alpha$}}
\epic
}

\def\axeaw{
\bpic(65,65)
  \put(5,5){\vector(1,1){55}}
  \put(55,55){\makebox(5,10){$\Omega$}}
  \put(55,5){\vector(-1,1){55}}
  \put(0,55){\makebox(5,10){$\alpha$}}
\epic
}

\def\axexyrhophi{
\bpic(65,65)
  \put(0,0){\axexy}
  \put(31,30){\vector(1,1){20}}
  \put(35,38){\makebox(5,5){$\rho$}}
  \put(35,30){\makebox(5,5){$\phi$}}
\epic
}

\def\axewxwyWt{
\bpic(65,65)
  \put(0,0){\axewxwy}
  \put(31,30){\vector(1,1){20}}
  \put(35,38){\makebox(5,5){$\Omega$}}
  \put(35,30){\makebox(5,5){$\theta$}}
\epic
}

\def\objet{
\bpic(40,40)
  \put(0,15){\line(0,1){10}}
  \put(0,25){\line(1,2){5}}
  \put(5,35){\line(1,1){5}}
  \put(10,40){\line(1,0){15}}
  \put(25,40){\line(1,-1){10}}
  \put(35,30){\line(0,-1){15}}
  \put(35,15){\line(-1,-1){10}}
  \put(25,5){\line(-1,0){10}}
  \put(15,5){\line(-1,1){5}}
  \put(10,10){\line(-2,1){10}}
\epic
}

\def\proj{
\bpic(27,27)
  \put(0,0){\line(1,2){2}}
  \put(2,4){\line(0,1){6}}
  \put(2,10){\line(1,1){4}}
  \put(6,14){\line(0,1){5}}
  \put(6,19){\line(1,1){4}}
  \put(10,23){\line(1,0){6}}
  \put(16,23){\line(1,1){4}}
  \put(20,27){\line(1,0){7}}
  \put(0,0){\line(1,1){27}}
\epic
}

\def\SystemLineaire{
\bpic(100,50)
  \put(5,15){\makebox(15,0){$f(x,y)$}}
  \put(20,15){\vector(1,0){10}}
  \put(30,7){\framebox(30,15){$H$}}
  \put(60,15){\vector(1,0){10}}
  \put(75,15){\circle{10}}
  \put(75,15){\makebox(0,0){$+$}}
  \put(80,15){\vector(1,0){10}}
  \put(90,15){\makebox(15,0){$g(x,y)$}}
  \put(75,30){\vector(0,-1){10}}
  \put(68,32){\makebox(15,0){$b(x,y)$}}
\epic
}

\def\grille#1{
\begin{picture}(50,50)
   \linethickness{0.01mm}
   \multiput(0,0)(5,0){#1}{\line(0,1){50}}
   \multiput(0,0)(0,5){#1}{\line(1,0){50}}
\end{picture}
}

\def\grillep#1{
\begin{picture}(50,50)
   \multiput(0,0)(0,5){#1}{\begin{picture}(0,0)\multiput(0,0)(5,0){#1}{\makebox(0,0){$.$}}\end{picture}}
   \multiput(0,0)(5,0){#1}{\begin{picture}(0,0)\multiput(0,0)(0,5){#1}{\makebox(0,0){$.$}}\end{picture}}
\end{picture}
}

\def\griller#1{
\begin{picture}(75,75)
   \linethickness{0.01mm}
   \multiput(25,0)(-3.47,3.47){#1}{\line(1,1){34.7}}
   \multiput(25,0)(3.47,3.47){#1}{\line(-1,1){34.7}}
\end{picture}
}

\def\slice{
\begin{picture}(140,70)
  \put(13,35){\makebox(20,15){$f(x,y)$}}
  \put(35,28){\makebox(5,10){$\phi$}}
  \put(80,10){\makebox(5,10){$p(r,\phi)$--FT--$P(\Omega,\phi)$}}
  \put(0,0){\axexy}
  \put(0,0){\axers}
  \put(10,10){\objet}
  \put(35,28){\makebox(5,10){$\phi$}}
  \put(35,-5){\proj}
  \put(80,0){\axewxwy}
  \put(80,0){\axeaw}
  \put(90,30){\makebox(20,15){$F(\omega_{x}, \omega_{y})$}}
  \put(115,28){\makebox(5,10){$\phi$}}
  \put(96,15){\line(1,1){30}}
\end{picture}
}

\def\PST{
\begin{picture}(130,70)
  \put(0,0){\axexy}
  \put(0,0){\axers}
  \put(10,35){\makebox(20,15){$f(x,y)$}}
  \put(70,0){\axewxwy}
  \put(70,0){\axeab}
\end{picture}
}

\def\tomoXa{
\begin{picture}(70,70)
  \put(13,35){\makebox(20,15){$f(x,y)$}}
  \put(-1,0){\axexy}
  \put(10,10){\objet}
  \put(5,5){\vector(1,1){55}}
  \put(55,55){\makebox(5,10){$r$}}
  \put(35,28){\makebox(5,10){$\phi$}}
  \put(63,9){\makebox(20,15){$\bullet$ D}}
  \put(60,0){\makebox(15,10){$p(r,\phi)$}}
  \put(8,51){\makebox(15,15){S $\bullet$}}
  \put(40,-5){\proj}
\end{picture}
}

\def\tomoXb{
\begin{picture}(70,70)
  \put(12,35){\makebox(20,15){$f(x,y)$}}
  \put(-1,0){\axexy}
  \put(10,10){\objet}
  \put(5,5){\vector(1,1){55}}
  \put(55,55){\makebox(5,10){$r$}}
  \put(35,28){\makebox(5,10){$\phi$}}
  \put(40,32){\makebox(5,10){$L_{r,\phi}$}}
  \put(63,9){\makebox(20,15){$\bullet$ D\'etecteur}}
  \put(60,0){\makebox(15,10){projection $p(r,\phi)$}}
  \put(8,51){\makebox(15,15){Source $\bullet$}}
  \put(63,18){\line(-1,1){40}}
  \put(40,-5){\proj}
\end{picture}
}

\def\tomoXc{
\begin{picture}(70,70)
  \put(10,30){\makebox(20,15){$f(x,y)$}}
  \put(-1,0){\axexy}
  \put(-1,0){\axers}
  \put(10,10){\objet}
  \put(33,28){\makebox(5,10){$\phi$}}
  \put(45,32){\makebox(5,10){$L_{r,\phi}$}}
  \put(60,5){\makebox(15,15){$p(r,\phi)$}}
  \put(8,51){\makebox(15,15){Source $\bullet$}}
  \put(63,9){\makebox(20,15){$\bullet$ D\'etecteur}}
  \put(63,18){\line(-1,1){40}}
  \put(40,-5){\proj}
  \put(30,30){\line(1,2){7}}
  \put(30,44){\line(1,0){7}}
  \put(37,30){\line(0,1){14}}
  \put(30,35){\makebox(5,5){$\rho$}}
  \put(30,30){\makebox(5,5){$\theta$}}
  \put(35,22){\makebox(5,10){$x$}}
  \put(15,35){\makebox(20,15){$y$}}
\end{picture}
}

\def\trDD{
\begin{picture}(70,70)
  \put(0,0){\axexy}
  \put(0,0){\axers}
  \put(10,10){\objet}
  \put(10,30){\makebox(20,15){$f(x,y)$}}
  \put(5,25){\makebox(20,15){$D$}}
  \put(46,45){\vector(1,1){8}}
  \put(50,50){\makebox(5,10){$\xib$}}
  \put(46,45){\vector(-1,1){8}}
  \put(40,50){\makebox(5,10){$\xib^\bot$}}
  \put(35,28){\makebox(5,10){$\phi$}}
  \put(52,22){\line(-1,1){30}}
  \put(40,32){\makebox(5,10){$L_{r,\phi}$}}
\end{picture}
}

\def\trDDD{
\begin{picture}(70,70)
  \put(-1.1,0){\axexyz}
  \put(10,35){\makebox(20,15){$f(x,y,z)$}}
  \put(30,30){\line(1,1){12}}
  \put(43,43){\line(1,1){1.5}}
  \put(46,46){\line(1,1){1.5}}
  \put(48,48){\vector(1,1){15}}
  \put(55,55){\makebox(5,10){$r$}}
  \put(48,48){\vector(1,1){8}}
  \put(50,50){\makebox(5,10){$\xib$}}
  \put(32,32){\surface}
  \put(40,50){\makebox(5,10){$S$}}
  \multiput(45,45)(2,2){2}{\line(1,1){1}}
  \multiput(48,48)(0,-2){6}{\line(0,-1){1}}
  \put(48,35){\line(0,-1){14}}
  \put(30,30){\line(2,-1){18}}
\end{picture}
}

\def\TRTF{
\begin{picture}(120,50)
  \put(10,30){\makebox(15,15){$f(\xb)$}}
  \put(18,38){\circle{15}}
  \put(100,30){\makebox(15,15){$\widetilde{f}(r,\xib)$}}
  \put(110,38){\circle{18}}
  \put(55,4){\makebox(15,15){$\widehat{f}(\omegab)$}}
  \put(63,11){\circle{15}}
  \put(25,37){\makebox(75,10){${\cal R}$}}
  \put(25,38){\vector(1,0){75}}
  \put(25,12){\makebox(30,10){${\cal F}_n$}}
  \put(23,33){\vector(3,-2){32}}
  \put(28,22){\makebox(30,10){${\cal F}_n^{-1}$}}
  \put(55,14){\vector(-3,2){31}}
  \put(70,12){\makebox(30,10){${\cal F}_1$}}
  \put(102,33){\vector(-3,-2){32}}
  \put(70,22){\makebox(30,10){${\cal F}_1^{-1}$}}
  \put(70,14){\vector(3,2){31}}
\end{picture}
}

\def\TRTFD{
\begin{picture}(120,50)
  \put(10,30){\makebox(15,15){$f(x,y)$}}
  \put(18,38){\circle{15}}
  \put(100,30){\makebox(15,15){$p(r,\phi)$}}
  \put(110,38){\circle{18}}
  \put(54.5,4){\makebox(15,15){$\widehat{f}(\omega_x,\omega_y)$}}
  \put(62,11){\circle{25}}
  \put(25,37){\makebox(75,10){${\cal R}$}}
  \put(25,38){\vector(1,0){75}}
  \put(25,12){\makebox(30,10){${\cal F}_2$}}
  \put(23,33){\vector(3,-2){32}}
  \put(28,22){\makebox(30,10){${\cal F}_2^{-1}$}}
  \put(55,14){\vector(-3,2){31}}
  \put(70,12){\makebox(30,10){${\cal F}_1$}}
  \put(102,33){\vector(-3,-2){32}}
  \put(70,22){\makebox(30,10){${\cal F}_1^{-1}$}}
  \put(70,14){\vector(3,2){31}}
\end{picture}
}

\def\TRTFs{
\begin{picture}(120,50)
  \put(10,30){\makebox(15,15){$f(\xb)$}}
  \put(18,38){\circle{15}}
  \put(109.5,38){\oval(16,14)}
  \put(109,38){\oval(16,14)}
  \put(100,30){\makebox(16,15){$\widetilde{f}(r,\xib)$}}
  \put(55,4){\framebox(15,15){\small $\barr{c} \widehat{f}(\omegab)\\ \omegab=\Omega\xib\earr$}}
  \put(25,37){\makebox(75,10){${\cal R}$}}
  \put(25,38){\vector(1,0){75}}
  \put(25,12){\makebox(30,10){${\cal F}_n$}}
  \put(23,33){\vector(3,-2){32}}
  \put(28,22){\makebox(30,10){${\cal F}_n^{-1}$}}
  \put(55,14){\vector(-3,2){31}}
  \put(70,12){\makebox(30,10){${\cal F}_1$}}
  \put(102,33){\vector(-3,-2){32}}
  \put(70,22){\makebox(30,10){${\cal F}_1^{-1}$}}
  \put(70,14){\vector(3,2){31}}
\end{picture}
}

\def\TRTFDs{
\begin{picture}(120,50)
  \put(10,30){\makebox(15,15){$f(x,y)$}}
  \put(18,38){\circle{15}}
  \put(109.5,38){\oval(16,14)}
  \put(109,38){\oval(16,14)}
  \put(100,30){\makebox(16,15){$p(r,\phi)$}}
  \put(48,4){\framebox(28,18){{\small 
   $\barr{c} \widehat{f}(\omega_x,\omega_y)=\widehat{p}(\Omega,\phi)\\ 
             \omega_x=\Omega\cosf\\ \omega_y=\Omega\sinf\earr$}}}
  \put(25,37){\makebox(75,10){${\cal R}$}}
  \put(25,38){\vector(1,0){75}}
  \put(25,12){\makebox(30,10){${\cal F}_2$}}
  \put(23,33){\vector(3,-2){26}}
  \put(28,22){\makebox(30,10){${\cal F}_2^{-1}$}}
  \put(49,14){\vector(-3,2){26}}
  \put(70,12){\makebox(30,10){${\cal F}_1$}}
  \put(101,33){\vector(-3,-2){26}}
  \put(70,22){\makebox(30,10){${\cal F}_1^{-1}$}}
  \put(76,14){\vector(3,2){26}}
\end{picture}
}

\def\retroprojection{
\begin{picture}(120,50)
  \put(10,30){\makebox(15,15){$f(x,y)$}}
  \put(18,38){\circle{15}}
  \put(100,30){\makebox(15,15){$p(r,\phi)$}}
  \put(108,38){\circle{15}}
  \put(54,4){\makebox(15,15){$b(x,y)$}}
  \put(62,11){\circle{15}}
  \put(25,37){\makebox(75,10){${\cal R}$}}
  \put(25,38){\vector(1,0){75}}
  \put(25,12){\makebox(30,10){${\cal D}$}}
  \put(23,33){\vector(3,-2){32}}
  \put(25,22){\makebox(30,10){Convolution: $*\frac{1}{\rho}$}}
  \put(55,14){\vector(-3,2){31}}
  \put(70,12){\makebox(30,10){${\cal B}$}}
  \put(102,33){\vector(-3,-2){32}}
\end{picture}
}

\def\invFourierP{
\begin{picture}(120,65)
  \put(10,5){\framebox(20,15){$\fwxwy$}}
  \put(90,5){\framebox(20,15){$\widehat{p}(\Omega, \phi)$}}
  \put(10,45){\framebox(20,15){$f(x,y)$}}
  \put(90,45){\framebox(20,15){$p(r, \phi)$}}
  \put(30,50){\makebox(60,15){$\cal R$}}
  \put(30,53){\vector(1,0){60}}
  \put(10,20){\makebox(10,25){${\cal F}_2^{-1}$}}
  \put(20,20){\vector(0,1){25}}
  \put(100,20){\makebox(10,25){${\cal F}_1$}}
  \put(100,45){\vector(0,-1){25}}
  \put(30,20){\makebox(60,15){$\widehat{f}(\omega_x,\omega_y) = \widehat{p}(\Omega, \phi)$}}
  \put(30,10){\makebox(60,15){$\omega_x=\Omega \cosf\, , \,\omega_y=\Omega \sinf$}}
  \put(90,12){\vector(-1,0){60}}
\end{picture}
}

\def\invFourier{
\begin{picture}(120,65)
  \put(10,5){\framebox(20,15){$\fwxwy$}}
  \put(90,5){\framebox(20,15){$\fthwfi$}}
  \put(10,45){\framebox(20,15){$f(x,y)$}}
  \put(90,45){\framebox(20,15){$\ftrfi$}}
  \put(30,50){\makebox(60,15){$\cal R$}}
  \put(30,53){\vector(1,0){60}}
  \put(10,20){\makebox(10,25){${\cal F}_2^{-1}$}}
  \put(20,20){\vector(0,1){25}}
  \put(100,20){\makebox(10,25){${\cal F}_1$}}
  \put(100,45){\vector(0,-1){25}}
  \put(30,10){\makebox(60,15){$\omega_x=\Omega \cosf\, , \,\omega_y=\Omega \sinf$}}
  \put(90,12){\vector(-1,0){60}}
\end{picture}
}

\def\invRetroD{
\begin{picture}(120,85)
  \put(10,5){\framebox(20,15){ $f^{\dag}(r,\phi)$ } }
  \put(90,5){\framebox(20,15){$\fthwfi$}}
  \put(45,5){\framebox(30,15){$\mid \Omega \mid \, \fthwfi$}}
  \put(10,65){\framebox(20,15){$f(x,y)$}}
  \put(90,65){\framebox(20,15){$\ftrfi$}}
  \put(45,35){\framebox(30,15){1--D Convolution}}
  \put(45,35){\vector(-1,-1){15}}
  \put(90,65){\vector(-1,-1){15}}
  \put(30,70){\makebox(60,15){$\cal R$}}
  \put(30,73){\vector(1,0){60}}
  \put(10,20){\makebox(5,45){${\cal B}$}}
  \put(20,20){\vector(0,1){45}}
  \put(100,20){\makebox(10,45){${\cal F}_1$}}
  \put(100,65){\vector(0,-1){45}}
  \put(30,10){\makebox(20,15){${\cal F}_1^{-1}$}}
  \put(45,12){\vector(-1,0){15}}
  \put(70,10){\makebox(20,15){$\mid \Omega \mid$}}
  \put(90,12){\vector(-1,0){15}}
\end{picture}
}

\def\invRetroDD{
\begin{picture}(120,85)
  \put(10,5){\framebox(20,15){ $f^{\dag}(r,\phi)$}}
  \put(90,5){\framebox(20,15){$b(x,y)$}}
  \put(45,5){\framebox(30,15){$\widetilde{b}(\omega_{x},\omega_{y})$}}
  \put(10,65){\framebox(20,15){$f(x,y)$}}
  \put(90,65){\framebox(20,15){$\ftrfi$}}
  \put(45,35){\framebox(30,15){2--D Convolution}}
  \put(100,65){\vector(0,-1){45}}
  \put(45,50){\vector(-1,1){15}}
  \put(30,70){\makebox(60,15){$\cal R$}}
  \put(30,73){\vector(1,0){60}}
  \put(10,20){\makebox(5,45){${\cal F}_2^{-1}$}}
  \put(20,20){\vector(0,1){45}}
  \put(100,20){\makebox(10,45){${\cal B}$}}
  \put(90,12){\vector(-1,0){15}}
  \put(30,20){\makebox(20,15){$\mid\Omega\mid=\sqrt{\omega_x^2+\omega_y^2}$}}
  \put(45,12){\vector(-1,0){15}}
  \put(75,10){\makebox(20,15){${\cal F}_2$}}
  \put(90,20){\vector(-1,1){15}}
\end{picture}
}

\def\migration{
\begin{picture}(70,70)
  \put(20,40){\makebox(20,15){Propagating}}
  \put(20,35){\makebox(20,15){region}}
  \put(35,25){\makebox(20,15){Evanecent}}
  \put(35,20){\makebox(20,15){region}}
  \put(-1,0){\axewxwy}
  \put(5,5){\framebox(50,50){}}
  \put(5,5){\line(1,1){50}}
  \put(55,5){\line(-1,1){50}}
\end{picture}
}

\def\DiffTomGDDD{
\begin{picture}(70,70)
  \put(0,0){\axexyz}
  \put(10,10){\objet}
  \put(12,35){\makebox(20,15){$f(\rb)$}}
  \put(30,30){\makebox(20,15){$\rb$}}
  \put(25,35){\makebox(20,15){$\rb '$}}
  \put(0,35){\makebox(10,15){$\psi_0(\rb)$}}
  \put(20,45){\makebox(20,15){$\psi(\rb)$}}
  \multiput(0,25)(2,0){3}{\line(0,1){10}}
  \put(0,30){\vector(1,0){10}}
\end{picture}
}

\def\DiffTomGDD{
\begin{picture}(70,70)
  \put(0,0){\axexy}
  \put(10,10){\objet}
  \put(12,35){\makebox(20,15){$f(x, y)$}}
  \put(30,30){\makebox(20,15){$\rb$}}
  \put(25,35){\makebox(20,15){$\rb '$}}
  \put(0,35){\makebox(10,15){$\psi_0(x)$}}
  \put(20,45){\makebox(20,15){$\psi(y)$}}
  \multiput(0,25)(2,0){3}{\line(0,1){10}}
  \put(0,30){\vector(1,0){10}}
\end{picture}
}

\def\interpSD{
\begin{picture}(125,120)
 \put(0,0){
  \begin{picture}(95,150)
  \put(0,50){\vector(1,0){125}}
  \put(120,45){\makebox(5,5){$x$}}
  \put(50,0){\vector(0,1){95}}
  \put(50,90){\makebox(5,5){$y$}}
  \put(10,10){\vector(1,1){95}}
  \put(105,105){\makebox(5,10){$x'$}}
  \put(95,5){\vector(-1,1){85}}
  \put(12,85){\makebox(5,10){$y'$}}
  \put(25,80){\makebox(5,10){$I_\phi(x', y')$}}
  \put(10,65){\makebox(5,10){$\phi=0$}}
  \put(10,55){\makebox(5,10){$\phi=45$}}
  \put(110,90){\makebox(5,10){$p(r,\phi)$}}
  \put(110,80){\makebox(5,10){$\phi=45$}}
  \put(115,70){\makebox(5,10){$\phi=0$}}
 \end{picture}}
 \put(25.1,25){\grille{11}}
 \put(25.1,15.2){\griller{11}}
 \put(110,25){\special{picture projv}}
 \put(75,75){\special{picture projo}}
\end{picture}
}

\def\interpFD{
\begin{picture}(65,80)
 \put(20,20){\axewxwy}
 \put(10,60){\makebox(5,15){$\widehat{f}(\Omega, \phi)$}}
 \put(10,75){\makebox(5,10){$\fwxwy$}}
 \put(25,25){\grille{11}}
 \put(25,25){\special{picture dcercles}}
\end{picture}
}

\def\analyse{
\begin{picture}(90,30)
  \put(5,15){\makebox(15,0){Objet}}
  \put(20,15){\vector(1,0){10}}
  \put(30,7){\framebox(30,15){Mod\`ele}}
  \put(60,15){\vector(1,0){10}}
  \put(72,15){\makebox(15,0){Projection}}
\end{picture}
}

\def\identification{
\begin{picture}(100,40)
  \put(0,25){\makebox(15,0){Objet}}
  \put(15,25){\vector(1,0){10}}
  \put(25,17){\framebox(30,15){Mod\`ele}}
  \put(55,25){\vector(1,0){10}}
  \put(67,25){\makebox(15,0){Projection}}
  \put(85,25){\vector(1,0){10}}
  \put(100,25){\circle{10}}
  \put(100,25){\makebox(0,0){$-$}}
  \put(92,42){\makebox(15,0){Mesures}}
  \put(100,40){\vector(0,-1){10}}
  \put(100,20){\line(0,-1){13}}
  \put(100,7){\vector(-1,0){15}}
  \put(55,0){\framebox(30,15){Identification}}
  \put(55,7){\line(-1,0){20}}
  \put(35,7){\vector(1,1){10}}
\end{picture}
}

\def\inversion{
\begin{picture}(100,40)
  \put(5,35){\makebox(15,0){Mesures}}
  \put(20,35){\vector(1,0){10}}
  \put(35,35){\circle{10}}
  \put(35,35){\makebox(0,0){$-$}}
  \put(40,35){\vector(1,0){10}}
  \put(50,27){\framebox(30,15){Inversion}}
  \put(80,35){\vector(1,0){20}}
  \put(80,40){\makebox(25,0){objet estim\'ee}}
  \put(90,35){\line(0,-1){20}}
  \put(90,15){\vector(-1,0){10}}
  \put(50,7){\framebox(30,15){Mod\`ele}}
  \put(50,15){\line(-1,0){15}}
  \put(35,15){\vector(0,1){15}}
\end{picture}
}

\def\syslin{
\begin{picture}(100,40)
  \put(5,15){\makebox(15,0){$f(t)$}}
  \put(20,15){\vector(1,0){10}}
  \put(30,7){\framebox(30,15){$H$}}
  \put(60,15){\vector(1,0){10}}
  \put(75,15){\circle{10}}
  \put(75,15){\makebox(0,0){$+$}}
  \put(75,30){\vector(0,-1){10}}
  \put(68,32){\makebox(15,0){$b(t)$}}
  \put(80,15){\vector(1,0){10}}
  \put(90,15){\makebox(15,0){$g(t)$}}
\end{picture}
}

\def\inversion{
\begin{picture}(100,40)
  \put(5,15){\makebox(15,0){$g(t)$}}
  \put(20,15){\vector(1,0){10}}
  \put(30,7){\framebox(30,15){$?$}}
  \put(60,15){\vector(1,0){10}}
  \put(70,15){\makebox(15,0){$\wh{f}(t)$}}
\end{picture}
}

\def\syslins#1#2#3#4{
\begin{picture}(100,40)
  \put(5,15){\makebox(15,0){#1}}
  \put(20,15){\vector(1,0){10}}
  \put(30,7){\framebox(30,15){#2}}
  \put(60,15){\vector(1,0){10}}
  \put(75,15){\circle{10}}
  \put(75,15){\makebox(0,0){+}}
  \put(75,30){\vector(0,-1){10}}
  \put(68,32){\makebox(15,0){#3}}
  \put(80,15){\vector(1,0){10}}
  \put(90,15){\makebox(15,0){#4}}
\end{picture}
}

\def\inversions#1#2#3{
\begin{picture}(100,40)
  \put(5,15){\makebox(15,0){#1}}
  \put(20,15){\vector(1,0){10}}
  \put(30,7){\framebox(40,15){#2}}
  \put(70,15){\vector(1,0){10}}
  \put(80,15){\makebox(15,0){#3}}
\end{picture}
}

\title{Computed tomography image reconstruction from only two projections}
\name{Ali Mohammad-Djafari}
\address{Laboratoire des Signaux et Syst\`emes, \\ 
Unit\'e mixte de recherche 8506 (CNRS-Sup\'elec-UPS) \\  
Sup\'elec, Plateau de Moulon, 3 rue Joliot Curie, 91192 Gif-sur-Yvette, 
France.}

\begin{document}
\maketitle

\begin{abstract}
This paper concerns the image reconstruction from a few projections in Computed Tomography (CT). The main objective of this paper is to show that the problem is so ill posed that no classical method, such as analytical methods based on inverse Radon transform, nor the algebraic methods such as Least squares (LS) or regularization theory can give satisfactory result. 
As an example, we consider in detail the case of image reconstruction from two horizontal and vertical projections. We then show how a particular composite Markov modeling and the Bayesian estimation framework can possibly propose 
satisfactory solutions to the problem. 
For demonstration and educational purpose a set of Matlab programs are 
given for a live presentation of the results. 
\end{abstract}

\section{Introduction}
The image reconstruction problem in Computed Tomography (CT) is presented in Fig.~1 and Fig.~2 shows the forward modeling of the problem via the Radon  Transform (RT) and the basics of the analytical RT inversion based methods such as Filtered backprojection \cite{Herman73,Hanson80,Herman80,Grangeat91,Natterer97,Djafari99a,Natterer99,djafari01z}. 

\medskip\noindent\fbox{\noindent\hbox{\vbox{
\bcc
\btabu{@{}cc@{}}
3D & 2D \\ 
\includegraphics[width=40mm,height=30mm]{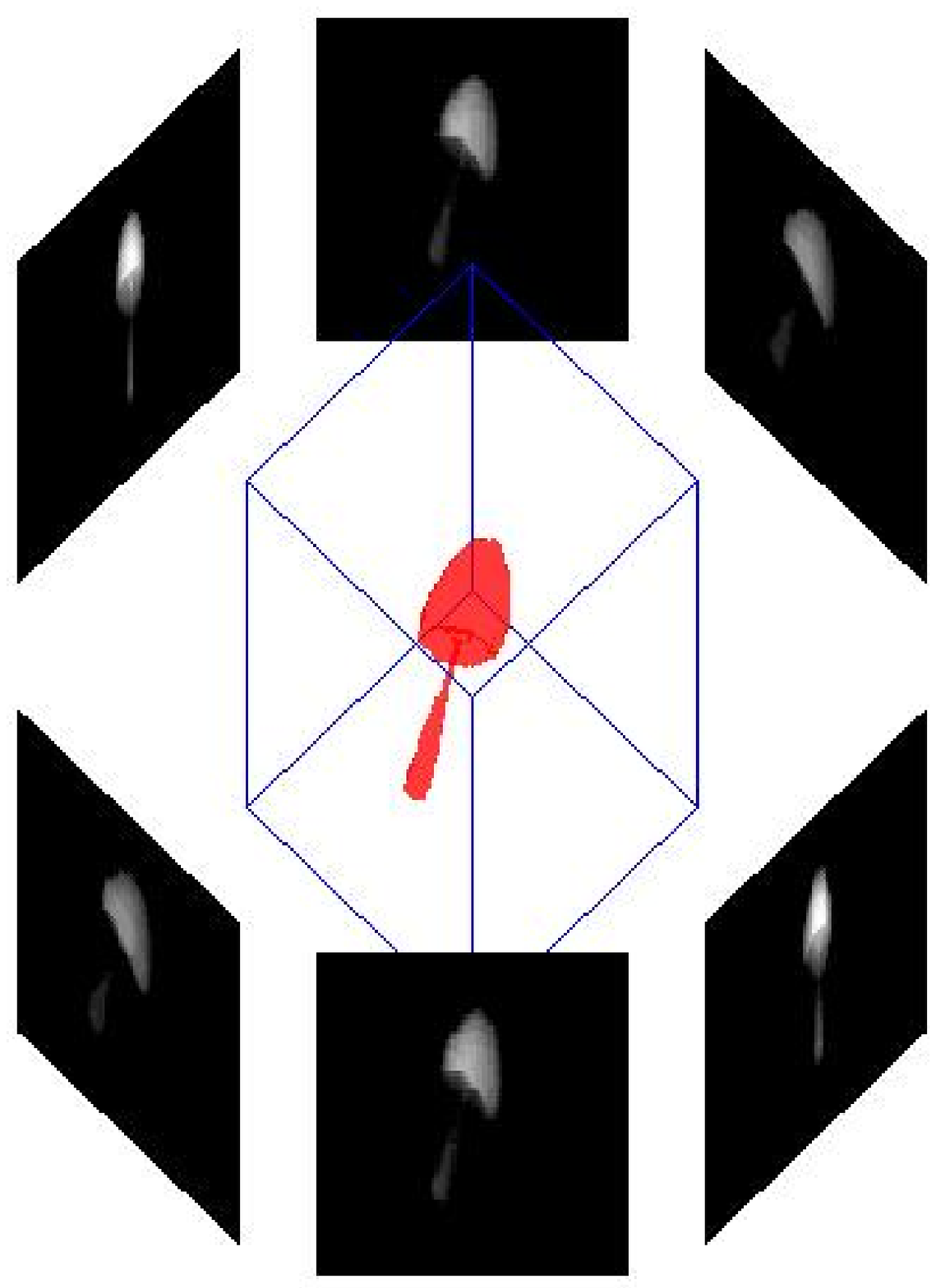} & 
\includegraphics[width=40mm,height=30mm]{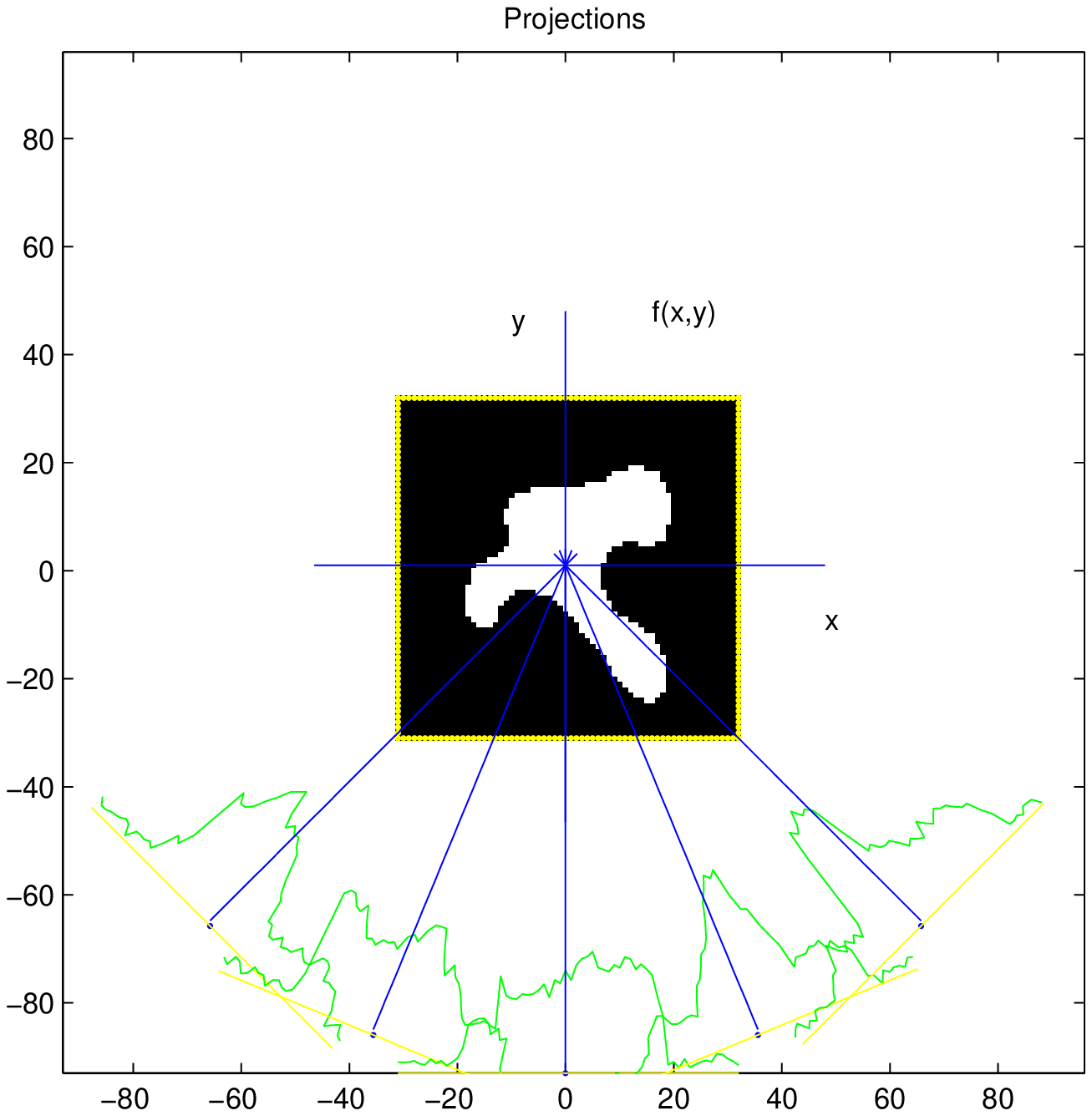} 
\etabu
\\
$\displaystyle{\blue{g_{\phi}(r_1,r_2)}=\int_{\Lc_{r_1,r_2,\phi}} \red{f(x,y,z)} \d{l}}$ \quad 
$\displaystyle{\blue{g_{\phi}(r)}=\int_{\Lc_{r,\phi}} \red{f(x,y)} \d{l}}$ 
\\ 
Forward problem:  
$\red{f(x,y)}$ or $\red{f(x,y,z)}\lra$ 
$\blue{g_{\phi}(r)}$ or $\blue{g_{\phi}(r_1,r_2)}$
\\
Inverse problem:  
$\blue{g_{\phi}(r)}$ or $\blue{g_{\phi}(r_1,r_2)}\lra$
$\red{f(x,y)}$ or $\red{f(x,y,z)}$  
\\ ~\\
{\bf Fig.~1~: Tomography X}
\ecc
}}}
\\ 

However, it is so evident that these methods cannot give satisfactory results 
in cases of very limited number of projections as is the case we consider in this papaer. To be able to introduce the necessary prior information needed to overcome the lack of information in the data, we consider the algebraic methods. Fig.~3 shows the discretization step of the forward problem which transforms the linear continuous RT equation to a system of finite linear equations which is  $\gb=\Hb\fb$. It is then evident that this system is under-determined and that the problem has an infinite number of solutions. 
As a demonstrative example, we consider the case of image reconstruction from only two projections and study the structure of the matrix $\Hb$ in this particular case and show easily that neither the neither the minimum norme least squares (MNLS) nor the generalized inversion and nor the quadratic regularization \cite{Demoment01} can give satisfactory result to this problem. 

\medskip\noindent\fbox{\noindent\hbox{\vbox{
\bcc
\vspace*{-5mm}
\thicklines\setlength{\unitlength}{.72mm}
\tomoXaz
\\ ~\\
\(
\barr{@{}c@{}}
\disp{\blue{g(r,\phi)}=\intd_D \red{f(x,y)} \, \delta(r-x\cos\phi-y\sin\phi) \d{x} \d{y}}
\\
\disp{\red{f(x,y)}= \left(-\frac{1}{2\pi^2} \right)
 \izpi\iii\frac{\dpdx{}{r} \blue{g(r,\phi)}}{(r-x\cos\phi-y\sin\phi)} \d{r}\d{\phi}}
\earr
\)
\[
\barr{@{}l@{}}
\mbox{Derivation $\Dc$:~~~~~~} 
\disp{\overline{g}(r,\phi)=\dpdx{g(r,\phi)}{r}} 
\\ 
\mbox{Hilbert Transform$\Hc$:~~} 
\disp{g_1(r',\phi)
=\frac{1}{\pi} \izi \frac{\overline{g}(r,\phi)}{(r-r')} \d{r}}
\\ 
\mbox{Backprojection $\Bc$:}
\disp{f(x,y)= \frac{1}{2\pi} \izpi g_1(x\cos\phi+y\sin\phi,\phi) \d{\phi}}
\earr 
\]
\[
\barr{@{}c@{}}
\disp{\red{f(x,y)} = \Bc \; \Hc \, \Dc \, \blue{g(r,\phi)}
= \Bc \; \Fc_1^{-1} \, |\Omega| \, \Fc_1 \, \blue{g(r,\phi)}}
\\ 
\stackrel{\blue{g(r,\phi)}}{\lra} 
\fbox{\btabu{@{}c@{}} {\small \textsc{FT}}\\${\cal F}_1$\etabu}\lra
\fbox{\btabu{@{}c@{}} {\small Filter}\\$|\Omega|$\etabu}\lra
\fbox{\btabu{@{}c@{}} {\small \textsc{IFT}}\\${\cal F}_1^{-1}$\etabu}
\stackrel{\blue{g_1(r,\phi)}}{\lra} 
\fbox{\small\btabu{@{}c@{}} {\small Backprojection}\\ $\cal B$\etabu}
\stackrel{\red{f(x,y)}}{\lra}
\earr
\]
%{\bf Fig.~2~: Analytical Methods and Filtered Backprojection}
{\bf Fig.~2~: X ray Tomography and Radon Transform}
\ecc
}}}
\\ 

We also show that even applying the positivity constraint is not enough to obtain satisfactory results and that there is a need for more informative 
prior knowledge. Finally, as the main contribution of this work, we show that the Bayesian inference framework and the composite Markov modeling can possibly be of great help to develop new reconstruction methods whith possibly satisfactoy results. We consider in particular a composite and hierarchical Intensity-labels Markov modeling with a Gauss-Markov modeling for the intensity field and a hidden Pottz Markov field for the region labels and propose new reconstruction methods which can be applied in many imaging systems, and particularly, in Non Destructive Testing (NDT) imaging applications. 

\section{Discretization of the problem}
As we mentionned before, for demonstration purpose, we conside the particular case of image reconstruction from only two projections. Also, for the sake of simplicity, we give details about a very reduced case of a $(4\times 4)$ pixels image. 

\medskip\noindent\fbox{\noindent\hbox{\vbox{
\bcc
\cresta\crestb
\\
$\disp{\blue{g_i}=\sum_{j=1}^n H_{ij} \; \red{f_j} \qquad\qquad\qquad H_{ij}=\{0,1\} }$
\\
$\blue{\left[\barr{c}
g_1\\ g_2\\ \vdots\\ \vdots\\ g_8
\earr\right]}=
\left[\barr{c}
1000100010001000\\ 0100010001000100\\ 0010001000100010\\ 0001000100010001\\ 0000000000001111\\ 0000000011110000\\ 0000111100000000\\ 1111000000000000
\earr\right]
\red{\left[\barr{c}
f_1\\ f_2\\ \vdots\\ \vdots\\ \vdots\\ f_{16}
\earr\right]}$
\\ 
{\bf Fig.~3~: Discr\'etisation du probl\`eme}
\ecc
}}}
\\ 

If we note by the vector $\fb=[f_1,\cdots,f_{16}]^t$ the pixels values of the image $f(x,y)$ 
and by the vector  
$\gb=[g_1,\cdots,g_8]^t$ 
the values of its two projections $g(r,\phi)$ along the horizontal 
$\phi=0$ and vertical $\phi=90$ angels, and assuming 
$\Delta x=1, \Delta y=1, \Delta r=1$, the we have: 
\[
\begin{array}{c@{}c}
\begin{array}{|c|c|c|c|} \cline{1-4}
f_1 & f_5 & f_9    & f_{13} \\[4pt] \cline{1-4}
f_2 & f_6 & f_{10} & f_{14} \\[4pt] \cline{1-4}
f_3 & f_7 & f_{11} & f_{15} \\[4pt] \cline{1-4}
f_4 & f_8 & f_{12} & f_{16} \\[4pt] \cline{1-4}
\end{array}
&
\begin{array}{c} 
g_8 \\[4pt] g_7 \\[4pt] g_6 \\[4pt] g_5
\end{array}
\\[4pt]
\begin{array}{cccc} 
g_1 & g_2 & g_3 & g_4
\end{array}
\end{array}
\quad
\begin{array}{c@{}c}
\begin{array}{|c|c|c|c|} \cline{1-4}
f_{11} & f_{12} & f_{13} & f_{14} \\[4pt] \cline{1-4}
f_{21} & f_{22} & f_{23} & f_{24} \\[4pt] \cline{1-4}
f_{31} & f_{32} & f_{33} & f_{34} \\[4pt] \cline{1-4}
f_{41} & f_{42} & f_{43} & f_{44} \\[4pt] \cline{1-4}
\end{array}
&
\begin{array}{c} 
g_{24} \\[4pt] g_{23} \\[4pt] g_{22} \\[4pt] g_{21}
\end{array}
\\[4pt]
\begin{array}{cccc} 
g_{11} & g_{12} & g_{13} & g_{14}
\end{array}
\end{array}
\]
Noting also by  
$\gb_1=[g_1,\cdots,g_4]^t=[g_{11},\cdots,g_{14}]^t$,  $\gb_2=[g_5,\cdots,g_8]^t=[g_{21},\cdots,g_{24}]^t$ 
and the matrices $\Ab_1$, $\Ab_2$ and $\Ab$ such that:
\[
\gb_1=\Ab_1 \fb, \quad \gb_2=\Ab_2 \fb, \quad \gb=\Ab \fb=
\cro{\begin{array}{c} 
\Ab_1 \\ \cline{1-1}  \Ab_2
\end{array}} \fb
\]
\REM{\[
\Ab_1=\cro{\barr{c}
1111000000000000\\[4pt] 
0000111100000000\\[4pt] 
0000000011110000\\[4pt] 
0000000000001111\\[4pt] 
\earr
}, 
\Ab_2=\cro{\barr{c}
0001000100010001\\[4pt] 
0010001000100010\\[4pt] 
0100010001000100\\[4pt] 
1000100010001000
\earr
}
\]
}
Then, considering the image  
\[
f =\cro{\barr{cccc}
 0 & 0 & 0 & 0\\
 0 & 1 & 1 & 0\\ 
 0 & 1 & 1 & 0\\ 
 0 & 0 & 0 & 0
\earr
}
\]
the following Matlab code lines:

\noindent{\tt  
A1=[ 
1 1 1 1 0 0 0 0 0 0 0 0 0 0 0 0;\\ 
0 0 0 0 1 1 1 1 0 0 0 0 0 0 0 0;\\
0 0 0 0 0 0 0 0 1 1 1 1 0 0 0 0;\\
0 0 0 0 0 0 0 0 0 0 0 0 1 1 1 1];\\[6pt] 
A2=[
0 0 0 1 0 0 0 1 0 0 0 1 0 0 0 1;\\  
0 0 1 0 0 0 1 0 0 0 1 0 0 0 1 0;\\  
0 1 0 0 0 1 0 0 0 1 0 0 0 1 0 0;\\ 
1 0 0 0 1 0 0 0 1 0 0 0 1 0 0 0]; \\[6pt] 
A=[A1;A2];\\[6pt] 
f=[0 0 0 0;0 1 1 0;0 1 1 0;0 0 0 0]; \\[4pt] 
p=A*f(:);
}
\\ 
gives~: 
\[
\gb^t=[0 \, 2 \, 2 \, 0 \, 0 \, 2 \, 2 \, 0]
\]
Thus, we have modeled the forward problem. Now, we are going to consider the inverse problem which is given $\gb$ find $\fb$.

It is evident that this inverse problem is under-determined, and that it has an infinite number of solutions. Here are four examples:
\[
\cro{\barr{cccc}
 0 & 0 & 0 & 0\\  
 0 & 0 & 2 & 0\\  
 0 & 2 & 0 & 0\\  
 0 & 0 & 0 & 0
\earr
}
\quad 
\cro{\barr{cccc}
 0 & 0 & 0 & 0\\ 
 0 & 2 & 0 & 0\\  
 0 & 0 & 2 & 0\\  
 0 & 0 & 0 & 0
\earr
}
\]
\[
\cro{\barr{cccc}
 -.5 & 0 & 0 & .5\\[4pt] 
 1 & 2 & 0 & -1\\[4pt] 
 -1 & 0 & 2 & 1\\[4pt] 
 0.5 & 0 & 0 & -.5
\earr
}
\quad 
\cro{\barr{cccc}
 -.5 & 0 & 0 & .5\\[4pt] 
 0 & 2 & 0 & 0\\[4pt] 
 0 & 0 & 2 & 0\\[4pt] 
 .5 & 0 & 0 & -.5
\earr
}
\]

\section{Backprojection as the adjoint operator of RT and its equivalent matrix transposition}
Comparing the continuous RT and its corresponding discretization and the 
adjoint Backprojection operator given in Fig.~2, and its corresponding discretization, we see easily the equivalence of Backprojection and the transposition of the the matrix $\Ab$. Thus, $\fbh=\Ab^t\gb$ corresponds to an image obtained by backprojecting the projections and it is easy to see that:
$\Ab^t=\cro{\begin{array}{c|c} 
\Ab_1^t & \Ab_2^t
\end{array}}$
Thus, computing this solution is easy:

\bver 
fh=A'*p;reshape(fh,4,4)
\ever
 
\[
\fbh =\cro{\barr{cccc}
 0 & 2 & 2 & 0\\ 
 2 & 4 & 4 & 2\\  
 2 & 4 & 4 & 2\\ 
 0 & 2 & 2 & 0
\earr
}
\]
We may also note that $\fbh=\Ab^t\gb=\Ab_1^t\gb_1+\Ab_2^t\gb_2$ is the addition of two images 
\[
\widehat{f} =\cro{\barr{cccc}
 0 & 2 & 2 & 0\\ 
 0 & 2 & 2 & 0\\  
 0 & 2 & 2 & 0\\ 
 0 & 2 & 2 & 0
\earr
}
+\cro{\barr{cccc}
 0 & 0 & 0 & 0\\  
 2 & 2 & 2 & 2\\  
 2 & 2 & 2 & 2\\  
 0 & 0 & 0 & 0
\earr
}
\]
each being the backprojection of each projection. 
We may also remark that this solution is not very far 
from the result of the convolution of the original image 
with the following impluse response
\[
\cro{\barr{ccc}
 0 & 1 & 0\\  
 1 & 2 & 1\\  
 0 & 1 & 0
\earr
}\]

\section{LS, MNLS and Generalized Inversion}
Let consider the two symetric matrices $\Ab^t\Ab$ and $\Ab\Ab^t$:
\[
\Ab\Ab^t=\cro{\begin{array}{c|c} 
\Ab_1\Ab_1^t & \Ab_1\Ab_2^t \\[4pt] \cline{1-2}
\Ab_2\Ab_1^t & \Ab_2\Ab_2^t 
\end{array}} 
\]
\[
\Ab\Ab^t 
=\cro{\barr{c|c} 4\Ib & \oneb \\[4pt] \cline{1-2} \oneb & 4 \Ib \earr}
=\cro{\barr{c}
 4~0~0~0~1~1~1~1\\  
 0~4~0~0~1~1~1~1\\  
 0~0~4~0~1~1~1~1\\  
 0~0~0~4~1~1~1~1\\  
 1~1~1~1~4~0~0~0\\  
 1~1~1~1~0~4~0~0\\  
 1~1~1~1~0~0~4~0\\  
 1~1~1~1~0~0~0~4
\earr
}
\]
\[
\Ab^t\Ab=\cro{\begin{array}{c|c} 
\Ab_1^t\Ab_1 & \Ab_2^t\Ab_1 \\[4pt] \cline{1-2}
\Ab_1^t\Ab_2 & \Ab_2^t\Ab_2 
\end{array}} 
\]
\[
\Ab_1^t\Ab_1 = \Ab_2^t\Ab_2 
=\cro{\barr{c|c} 
\oneb+\Ib & \Ib \\[4pt] \cline{1-2} 
\Ib & \oneb+\Ib  
\earr}
\]
\[
\Ab_1^t\Ab_2 = \Ab_2^t\Ab_1 
=\cro{\barr{c|c} 
\Ib & \Ib \\[4pt] \cline{1-2} 
\Ib & \Ib 
\earr}, 
\]
\[
\Ab^t\Ab 
=\cro{\barr{c}
2~1~1~1~1~0~0~0~1~0~0~0~1~0~0~0\\ 
1~2~1~1~0~1~0~0~0~1~0~0~0~1~0~0\\ 
1~1~2~1~0~0~1~0~0~0~1~0~0~0~1~0\\ 
1~1~1~2~0~0~0~1~0~0~0~1~0~0~0~1\\ 
1~0~0~0~2~1~1~1~1~0~0~0~1~0~0~0\\ 
0~1~0~0~1~2~1~1~0~1~0~0~0~1~0~0\\ 
0~0~1~0~1~1~2~1~0~0~1~0~0~0~1~0\\ 
0~0~0~1~1~1~1~2~0~0~0~1~0~0~0~1\\ 
1~0~0~0~1~0~0~0~2~1~1~1~1~0~0~0\\ 
0~1~0~0~0~1~0~0~1~2~1~1~0~1~0~0\\ 
0~0~1~0~0~0~1~0~1~1~2~1~0~0~1~0\\ 
0~0~0~1~0~0~0~1~1~1~1~2~0~0~0~1\\ 
1~0~0~0~1~0~0~0~1~0~0~0~2~1~1~1\\ 
0~1~0~0~0~1~0~0~0~1~0~0~1~2~1~1\\ 
0~0~1~0~0~0~1~0~0~0~1~0~1~1~2~1\\ 
0~0~0~1~0~0~0~1~0~0~0~1~1~1~1~2
\earr
}
\]
and compute their singular values: 

\bver 
AAt=A*A'; svd(AAt)
\ever
 
\[
\mbox{svd}(\Ab\Ab^t)=[8 \, 4 \, 4 \, 4 \, 4 \, 4 \, 4 \, 0]
\]

\bver 
AtA=A'*A; svd(AtA);
\ever
 
\[
\mbox{svd}(\Ab^t\Ab)
=[8 \, 4 \, 4 \, 4 \, 4 \, 4 \, 4 \, 0 \, 0 \, 0 \, 0 \, 0 \, 0 \, 0 \, 0 \, 0]
\]
We can then remark that both are singular. 

We may remind that the least squares (LS) solutions are defined as 
\[
\fbh=\argmin{\fb}{\| \gb-\Ab\fb \|^2}, 
\]
and if $\Ab^t\Ab$ was invertible, then we had:  
$\fbh=(\Ab^t\Ab)^{-1}\Ab^t\gb$. 

In the same way, the minimum norme solution is defined as:
\[
\fbh=\argmin{\Ab\fb=\gb}{\|\fb \|^2} 
\]
and if $\Ab\Ab^t$ was invertible, then we had: 
$\fbh=\Ab^t(\Ab\Ab^t)^{-1}\gb$. 

As we noticed, the two matrices $\Ab\Ab^t$ and $\Ab^t\Ab$ are singular 
and thus we can not define those solutions. However, if we only consider their  diagonal elements, we can define: 
 
\bver 
fh=diag(1./diag(AtA))*A'*p;reshape(fh,4,4)
\ever
 
\[
\fbh =\cro{\barr{cccc}
 0 & 1 & 1 & 0\\[4pt] 
 1 & 2 & 2 & 1\\[4pt] 
 1 & 2 & 2 & 1\\[4pt] 
 0 & 1 & 1 & 0
\earr
}
\]
\bver 
fh=A'*diag(1./diag(AAt))*p;reshape(fh,4,4)
\ever 

\[
\fbh =\cro{\barr{cccc}
  0 & .5 & .5 & 0\\[4pt] 
 .5 & 1  & 1  & .5\\[4pt] 
 .5 & 1  & 1  & .5\\[4pt] 
  0 & .5 & .5 & 0
\earr
}
\]
Also, we can use the technic of truncation of the singular values 
to define an unique generalized inverse solution 
\[
\fbh=\sum_{k=1}^k \frac{<\gb,\ub_k>}{\lambda_k} \vb_k
\]
where $\ub_k$ and $\vb_k$ are, respectively, the eigenvectors of 
$\Ab\Ab^t$ and $\Ab^t\Ab$ and $\lambda_k$ their corresponding 
eigen values. 

\smallskip\noindent{\tt 
[U,S,V]=svd(A);\\   
s=diag(S);s1=[1./s(1:7);zeros(1,1)];\\ 
S1=[diag(s1);zeros(8,8)];\\
fh=V*S1*U'*p;reshape(fh,4,4)\\
} 

In this example, $K=7$ and the GI solution can be computed by

\bver 
fh=svdpca(A,p,.1,7);reshape(fh,4,4)
\ever 

\[
\fbh =\cro{\barr{cccc}
-0.2500 & 0.2500 & 0.2500 & -0.2500\\[4pt] 
 0.2500 & 0.7500 & 0.7500 &  0.2500\\[4pt] 
 0.2500 & 0.7500 & 0.7500 &  0.2500\\[4pt] 
-0.2500 & 0.2500 & 0.2500 & -0.2500                                      
\earr
}
\]
or by the following iterative algorithm:

\smallskip\noindent{\tt 
for k=1:100; \\ 
fh=fh+.1*A'*(p-A*fh(:)); \\ 
end;\\ 
reshape(fh,4,4)
}

\[
\fbh =\cro{\barr{cccc}
-0.2500 & 0.2500 & 0.2500 & -0.2500\\[4pt] 
 0.2500 & 0.7500 & 0.7500 &  0.2500\\[4pt] 
 0.2500 & 0.7500 & 0.7500 &  0.2500\\[4pt] 
-0.2500 & 0.2500 & 0.2500 & -0.2500                                      
\earr
}
\]

We may note that the Kernel of $\gb=\Ab\fb$, 
\ie~ $\acc{\fb | \Ab\fb=0}$ is given by 
\[
\Vb(\Ib-\Sb^+\Sb)\zb=\sum_{k=K+1}^N z_k\, \vb_k 
\]
where $\zb$ can be any arbitrary image. This can be used to obtain all the possible solutions of $\gb=\Ab\fb$ by adding these arbitrary images to the GI solution. 

\section{Regularisation}
We may easily note that $\Ab^t\Ab+\lambda\Ib$ and $\Ab\Ab^t+\lambda\Ib$ 
are no more singular if $\lambda>0$. Thus, we may compute:

\smallskip\noindent{\tt 
lambda=.01;\\ 
fh=inv(AtA+lambda*eye(size(AtA)))*(A'*p); \\ 
reshape(fh,4,4)
}

\[
\fbh =\cro{\barr{cccc}
-0.2491 & 0.2497 & 0.2497 & -0.2491\\[4pt] 
 0.2497 & 0.7484 & 0.7484 & 0.2497\\[4pt] 
 0.2497 & 0.7484 & 0.7484 & 0.2497\\[4pt] 
-0.2491 & 0.2497 & 0.2497 & -0.2491
\earr
}
\]
or still

\smallskip\noindent{\tt 
lambda=.01;\\  
fh=A'*inv(AAt+lambda*eye(size(AAt)))*p; \\  
reshape(fh,4,4)\\ 
}
\\[4pt] 
which give the same solution. 

\section{Positivity Constraint}
We may remark that the problem is so ill-posed that, imposing to the solutions to be of the minimum norme does not reduce enough the space of the possible solutions. The positivity constraint has been frequently used in many image reconstruction applications. A very simple technique to impose the positivity constraint in iterative algorithms is just to impose it at each iteration: 

\smallskip\noindent{\tt 
for k=1:100\\ 
fh=fh+.1*A'*(p-A*fh(:));\\ 
fh=fh.*(fh>0);\\ 
end\\ 
reshape(fh,4,4);
}

\[
fh =\cro{\barr{cccc} 
         0 &  0.0000  &  0.0000  &       0\\[4pt] 
    0.0000 &  1.0000  &  1.0000  &  0.0000\\[4pt] 
    0.0000 &  1.0000  &  1.0000  &  0.0000\\[4pt] 
         0 &  0.0000  &  0.0000  &       0                                       \earr
}
\]

Of course, this technique is only one of the possible methods to use 
the prior information of the positivity. However, we see that it can be very useful at least in this low scale case. But, as we will see later, in a real larger image reconstruction problem, it is not enough. 

\section{Real size images implementation of the algorithmes}
Let consider a $(256\times 256)$ pixel image. 
Then, it is no more question of really constructing the matrix $\Ab$, because its dimensions are $(256^2\times 256)$. Indeed, we do not really need its construction, we only need the results of the forward computation $\Ab\fb$ and the backprojection $\Ab^t\gb$.
The following programs shows how to compute these quantities without actually constructing the matrix $\Ab$. 

\smallskip\noindent{\tt 
function p=direct(f);\\ 
p1=sum(f);\\ 
p2=sum(f');\\ 
p=[p1(:);p2(:)];\\ 
return
}

\smallskip\noindent{\tt 
function f=transp(p);\\ 
l=length(p);p1=p(1:l/2);p2=p(l/2+1:l);\\ 
f=ones(l/2,1)*p1'+p2*ones(1,l/2);\\ 
return
}

These programs can easily be used to obtain the following results/

\noindent\fbox{\noindent\hbox{\vbox{
\bcc
\btabu{@{}c@{}c@{}}
a)\includegraphics[width=40mm,height=40mm]{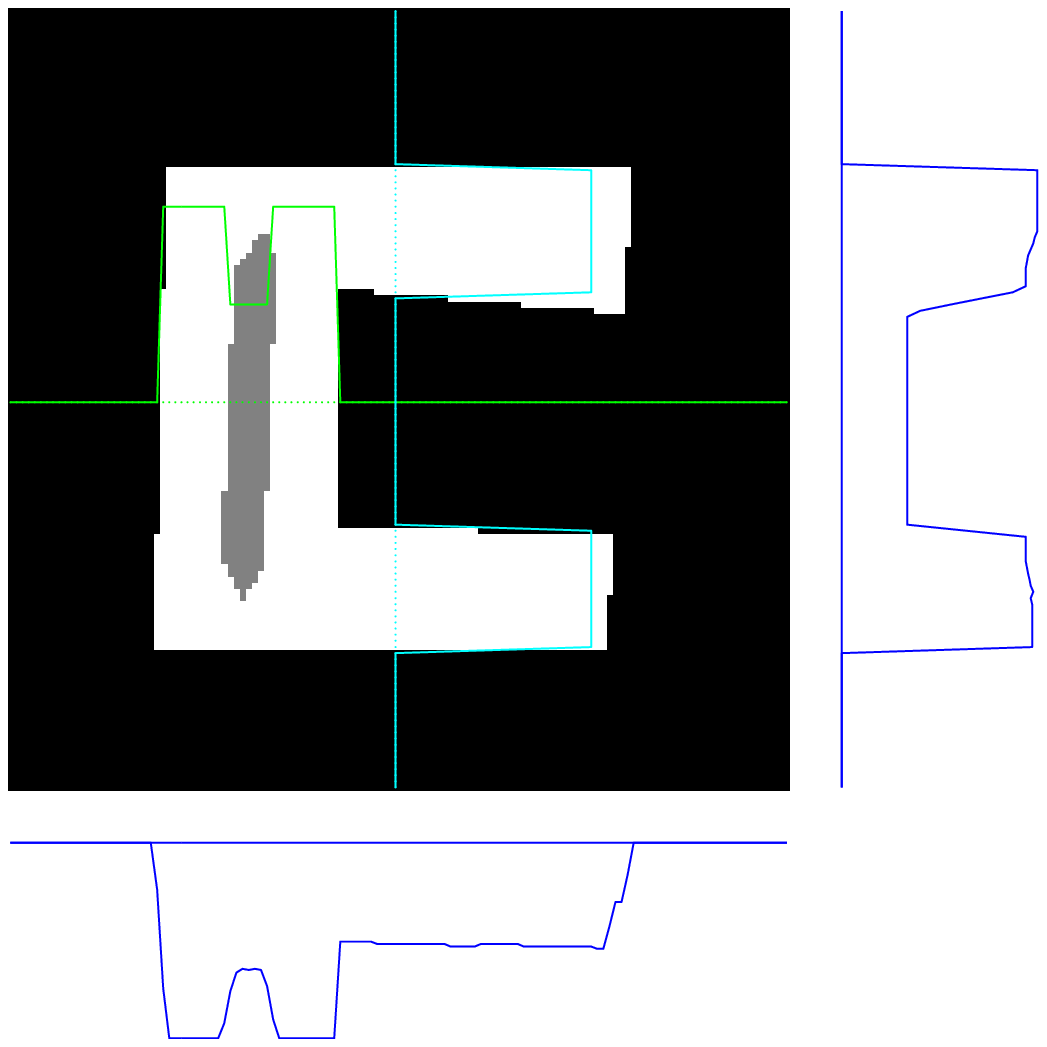} &
b)\includegraphics[width=40mm,height=40mm]{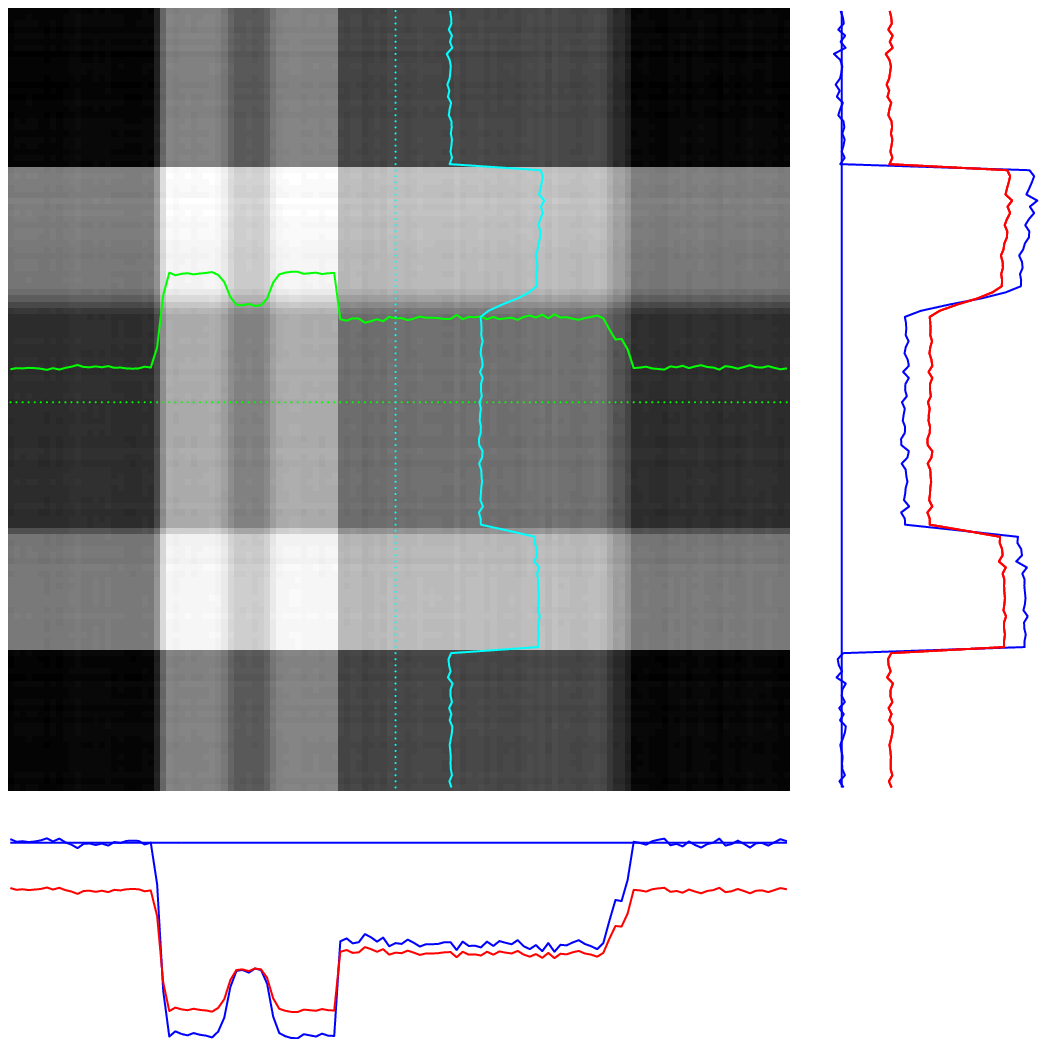}
\etabu
\\
{\bf Fig.~4~:} a) Objet et ses projections, 
b) R\'esultat de r\'etroprojection.
\ecc
}}}

We can also use them to impose any constraints such as positivity in the iterative methods:

\noindent{\bf Moindre carr\'e avec contraint de positivit\'e~:}

\smallskip\noindent{\tt 
alpha=.1;\\
for k=1:100\\
g=trans(p-direct(fh);\\ 
fh=fh+alpha*g;\\
fh=fh.*(fh>0);\\
end
}

\noindent{\bf R\'egularisation quadratique avec contraint de positivit\'e~:}

\smallskip\noindent{\tt 
alpha=.1;d=[-1 0 -1;0 4 0;-1 0 -1];\\
for k=1:100\\
g0=trans(p-direct(fh);\\ 
g=g0-lambda*conv2(fh,d,'same');\\ 
fh=fh+alpha*g\\
fh=fh.*(fh>0);\\
end
}

Remarquons que les algorithmes pr\'esent\'es plus haut sont assez 
rudimentaires (gradient \`a pas constant et \`a nombre d'it\'erations fini).
Il est \'evident que l'on peux faire mieux. \`A titre d'exemple, 
nous avons d\'evelopp\'e un logiciel d'optimisation (gpave) un peu 
plus \'elabor\'e qui 
met en {\oe}uvre d'autre algorithmes d'optimisation, comme par exemple, 
gradient \`a pas adaptative, gradient conjugu\'e et d'autres encore.

\noindent\fbox{\noindent\hbox{\vbox{
\bcc
\btabu{@{}c@{}c@{}}
a)\includegraphics[width=40mm,height=40mm]{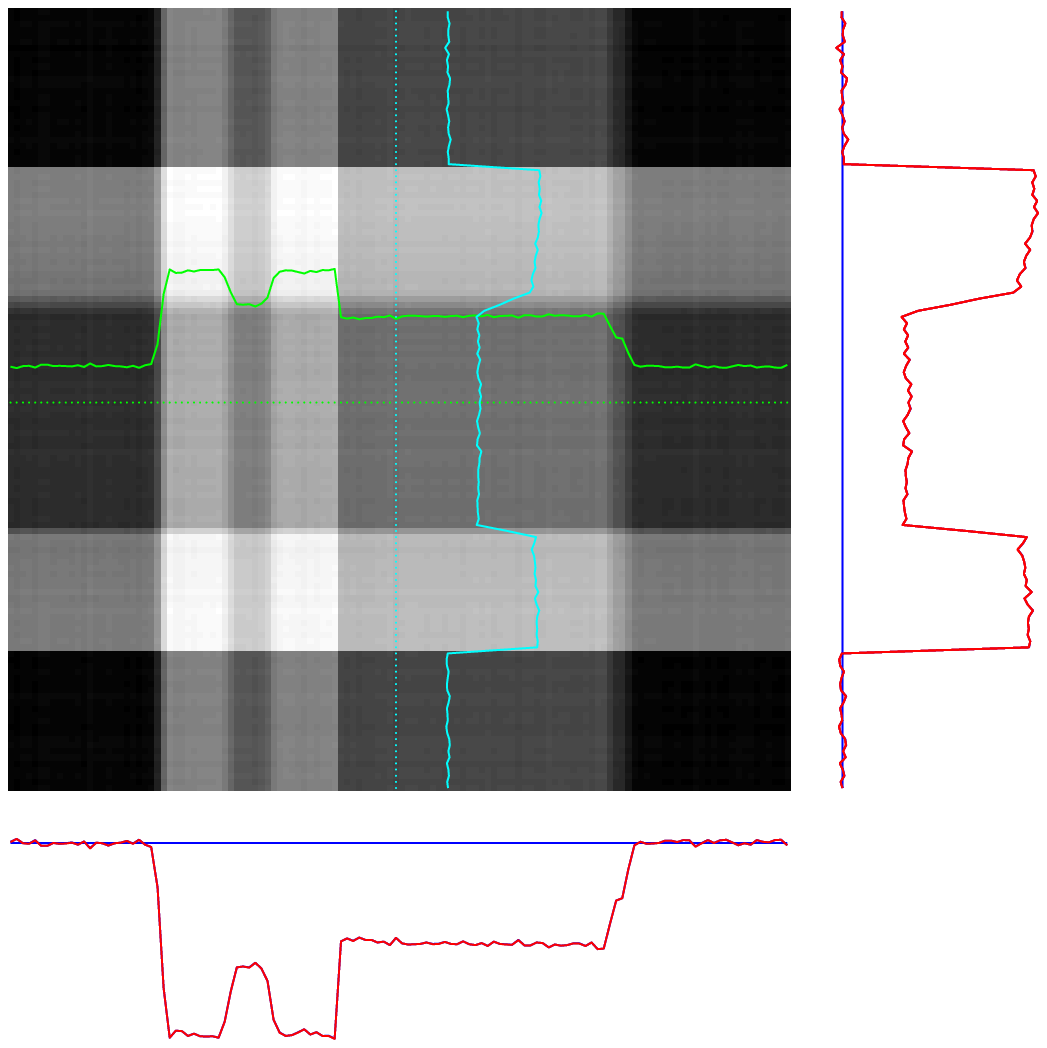} &
b)\includegraphics[width=40mm,height=40mm]{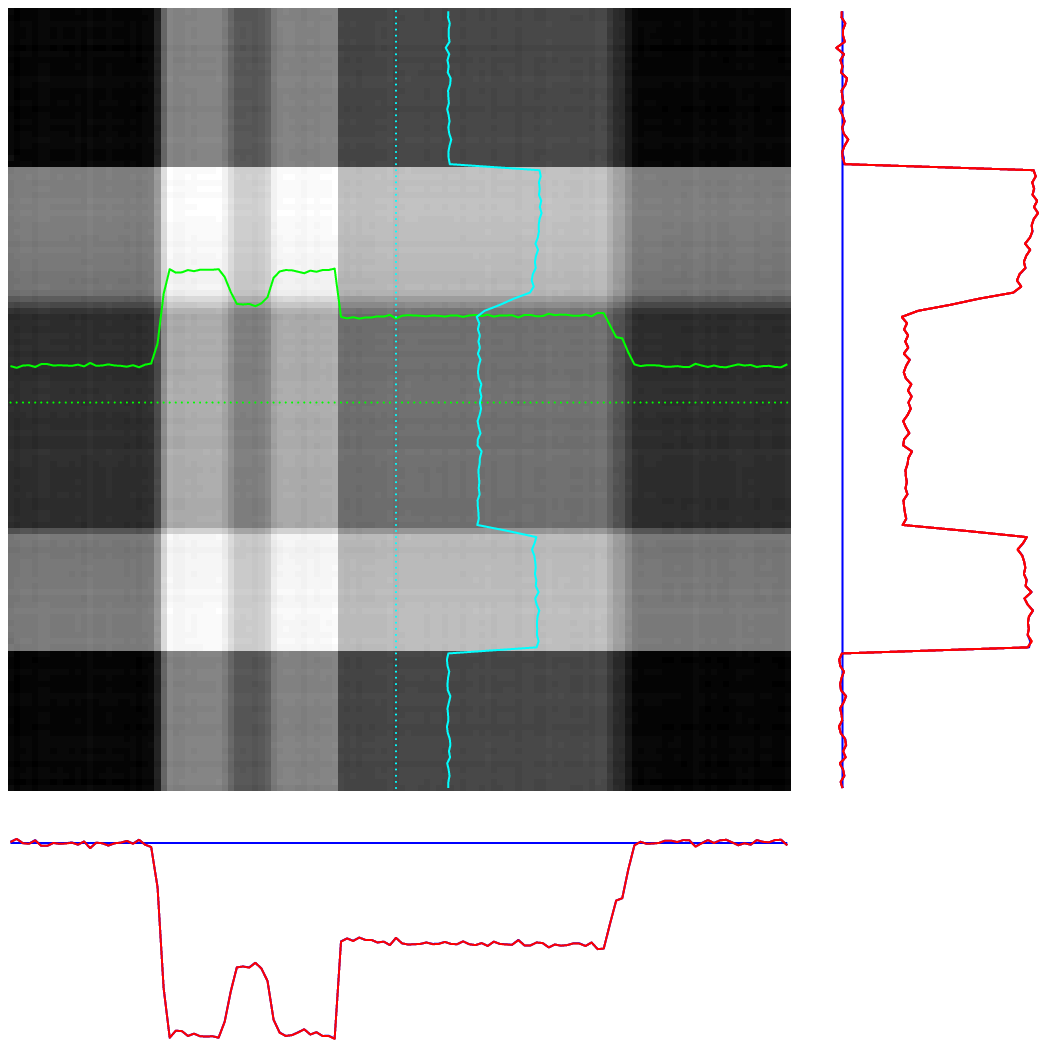}
\etabu
\\
{\bf Fig.~5~:} a) Moindre carr\'es avec contrainte de positivit\'ee et  
b) R\'egularisation quadratique avec contrainte de positivit\'ee.
\ecc
}}}

Les lignes de codes Matlab qui suivent montrent l'usage de ce logiciel.
Il faut tout d'abord \'ecrire deux routines qui calculent le crit\`ere 
{\tt crit} qui calcule 
\[
J=\|\gb-\Ab\fb\|^2+\lambda\|\Db\fb\|^2
\]
et son gradient {\tt dcrit} 
\[
\nabla J=-2\Ab^t(\gb-\Ab\fb)+2\lambda\Db^t\Db\fb
\]  
o\`u $\Db\fb$ correspond \`a l'application d'une op\'eration de convolution 
de l'image $f(i,j)$ avec une r\'eponse impulsionnelle 
$\left[\barr{cc} -1 & 1\\ 1 & -1\earr\right]$ ce qui correspond \`a  
\[
\sum_i\sum_j\left(|f(i,j)-f(i-1,j)|^2+|f(i,j)-f(i,j-1)|^2\right).
\] 
Notez aussi que $\Db^t\Db\fb$ correspond \`a l'application d'une op\'eration 
de convolution de l'image $f(i,j)$ avec une r\'eponse impulsionnelle 
\[
\left[\barr{ccc} -1 & 0 &-1\\ 0 & -4 & 0 \\ -1 & 0 &-1\earr\right]= 
\left[\barr{cc} -1 & 1\\ 1 & -1\earr\right]*\left[\barr{cc} -1 & 1\\ 1 & -1\earr\right]
\] 
o\`u $*$ signifie une convolution.

%+\lambda\sum_i\sum_j\left(|f(i,j)-f(i-1,j)|^2+|f(i,j)-f(i,j-1)|^2\right)$ 

\smallskip\noindent{\tt 
function J=crit(fh,p,lambda)\\
dp=p-direct(fh); \\
J0=sum(dp(:).\^2);\\
d=[-1 1;1 -1];\\
df=conv2(fh,d,'same');\\ 
J1=sum(df(:).\^2);\\
J=J0+lambda*J1;\\
return
}

\smallskip\noindent{\tt 
function dJ=dcrit(fh,p,lambda)\\
dp=p-direct(fh); \\
dJ0=-2*transp(dp);\\
d=[-1 0 -1;0 4 0;-1 0 -1];\\ 
dJ1=conv2(fh,d,'same');\\
dJ=dJ0+lambda*dJ1;\\
return
}

%\noindent 
Avec ces deux routines, le programme de la reconstruction devient tr\`es simple:

\smallskip\noindent{\tt 
f0=transp(p);\\ 
options = goptions;\\
lambda=1;\\ 
fh=gpav('crit',f0,options,'dcrit',p,lambda);\\
reshape(fh,4,4)\\
}

Avec ce programme d'optimisation il est alors facile de modifier 
les routines {\tt crit} et {\tt dcrit} pour changer le crit\`ere de 
la r\'egularisation. 
Les figures suivantes montrent un certain nombre des r\'esultats.

\noindent\fbox{\noindent\hbox{\vbox{
\bcc
\btabu{@{}c@{}c@{}}
\includegraphics[width=30mm,height=30mm]{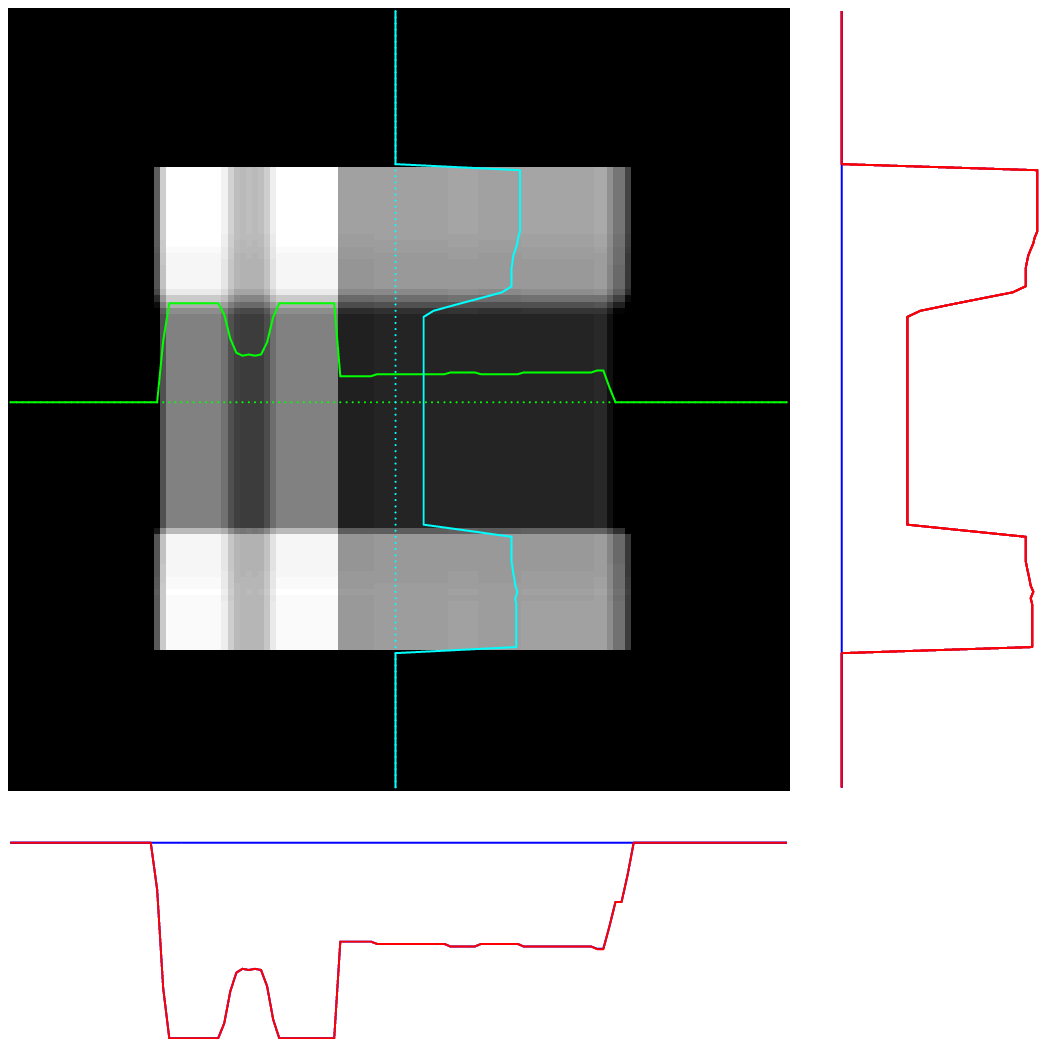}
\includegraphics[width=30mm,height=30mm]{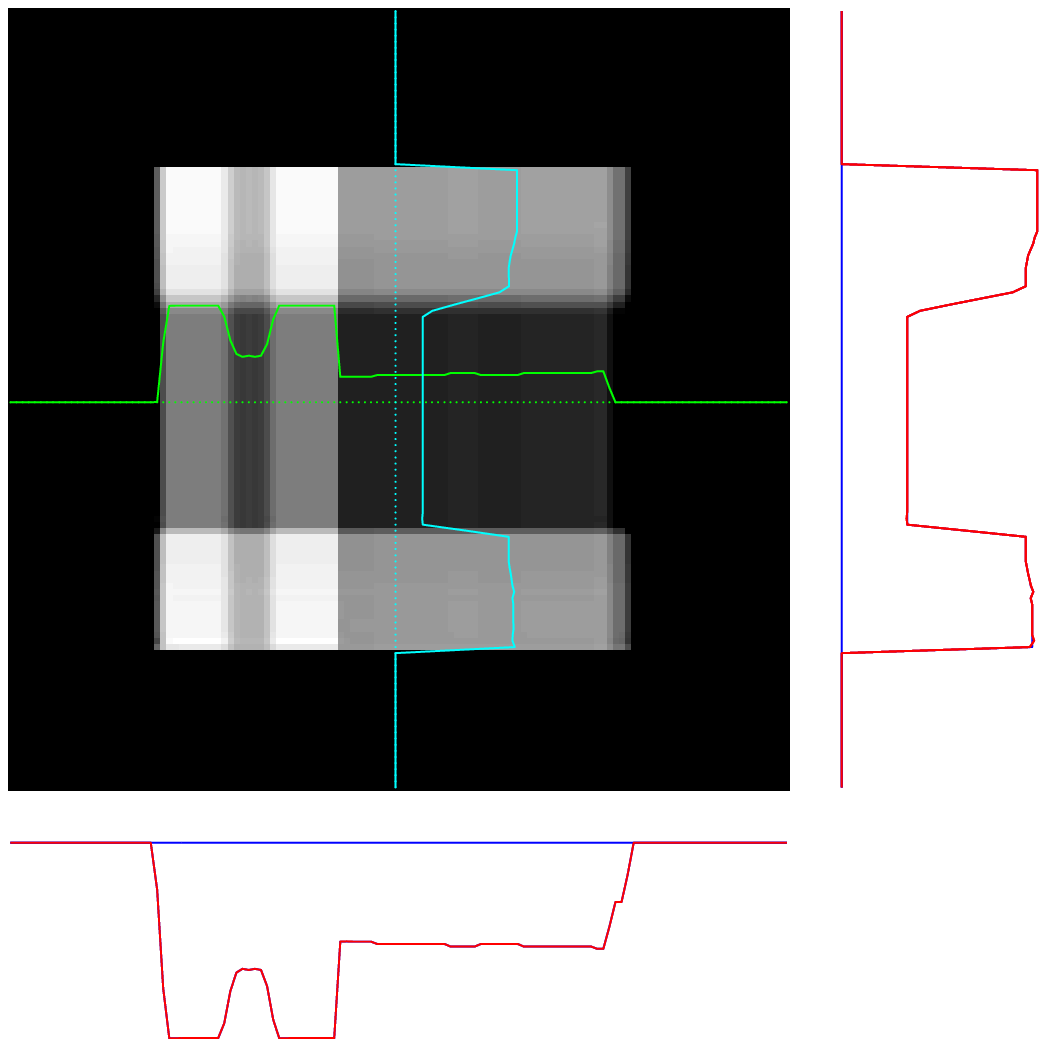}\\
a & b
\etabu
\\
{\bf Fig.~6~:} a) Moindre carr\'es avec contraintes de positivit\'ee 
et de support 
b) R\'egularisation quadratique avec contrainte de positivit\'ee et de support.  
\ecc
}}}

On remarque que le probl\`eme est tr\`es mal-conditionn\'ee au sens 
que la manque d'information dans les donn\'ees est trop important.
Il faut pouvoir obtenir d'autres donn\'ees, \ie, des projections 
suivant d'autre angles. 
Pour cela il faut r\'e\'ecrire les routines {\tt directe} et {\tt transp} 
et les routines correspondantes {\tt crit} et  {\tt dcrit} afin de pouvoir impl\'ementer le cas g\'en\'eral du calcul des projections suivant n'importe quel angle et la r\'etroprojection associ\'ee. 

Les figures suivantes montrent des exemples 
de reconstructions pour le cas o\`u on a 7 projections.

\noindent\fbox{\noindent\hbox{\vbox{
\bcc
\noindent\btabu{@{}c@{}c@{}c@{}}
a) \includegraphics[width=35mm,height=35mm]{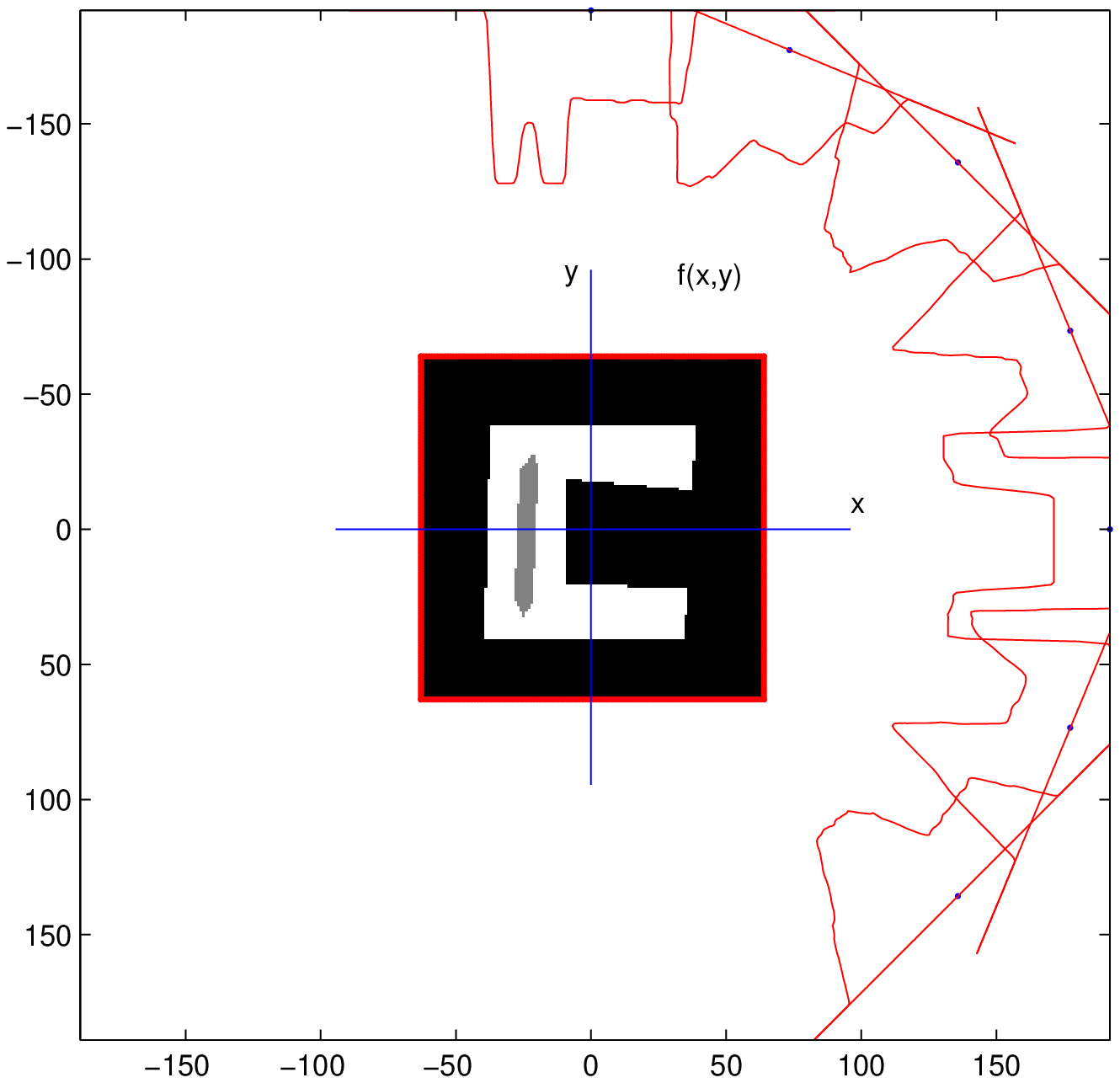} &
b) \includegraphics[width=35mm,height=35mm]{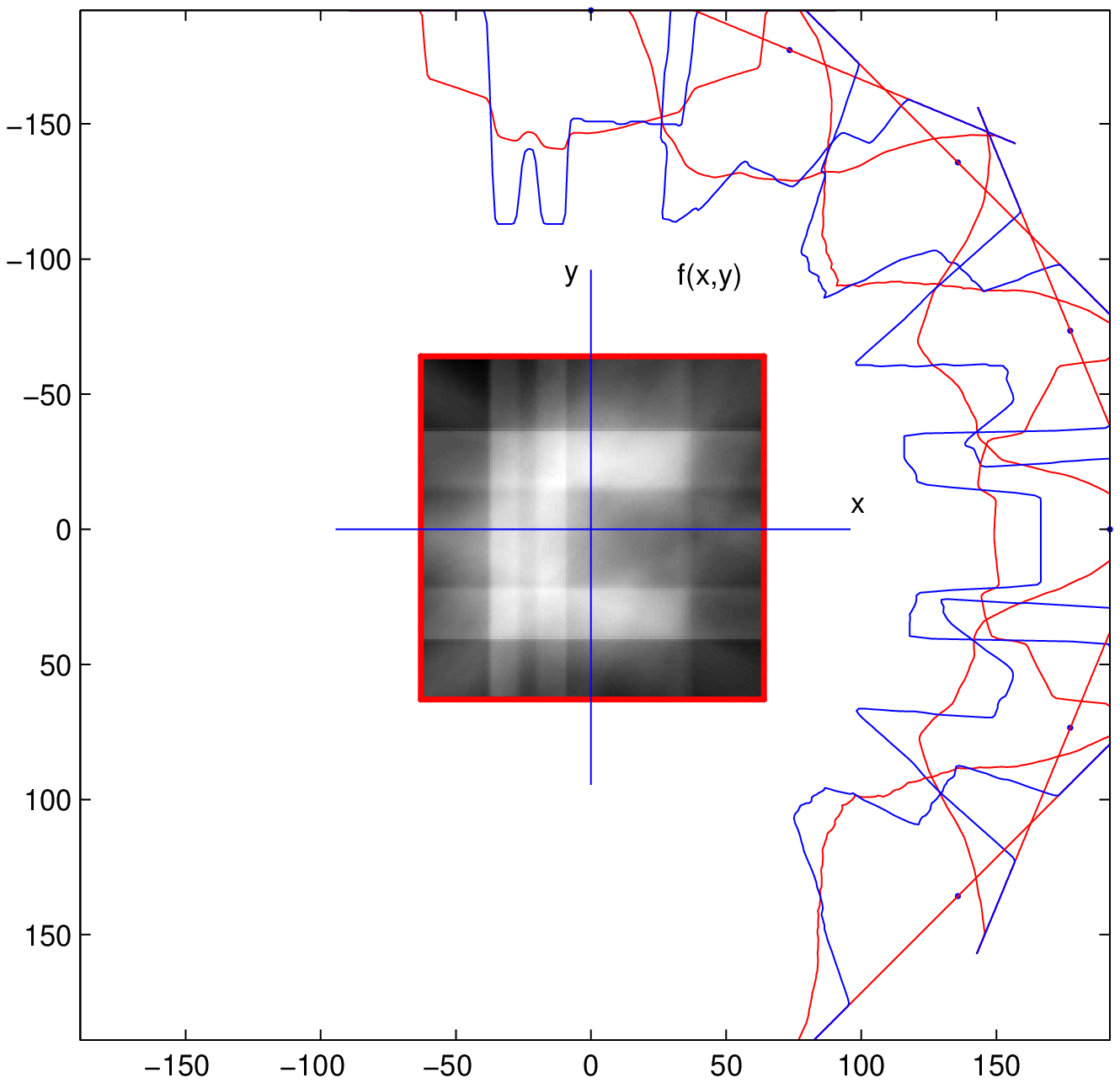}\\[6pt]  
c) \includegraphics[width=35mm,height=35mm]{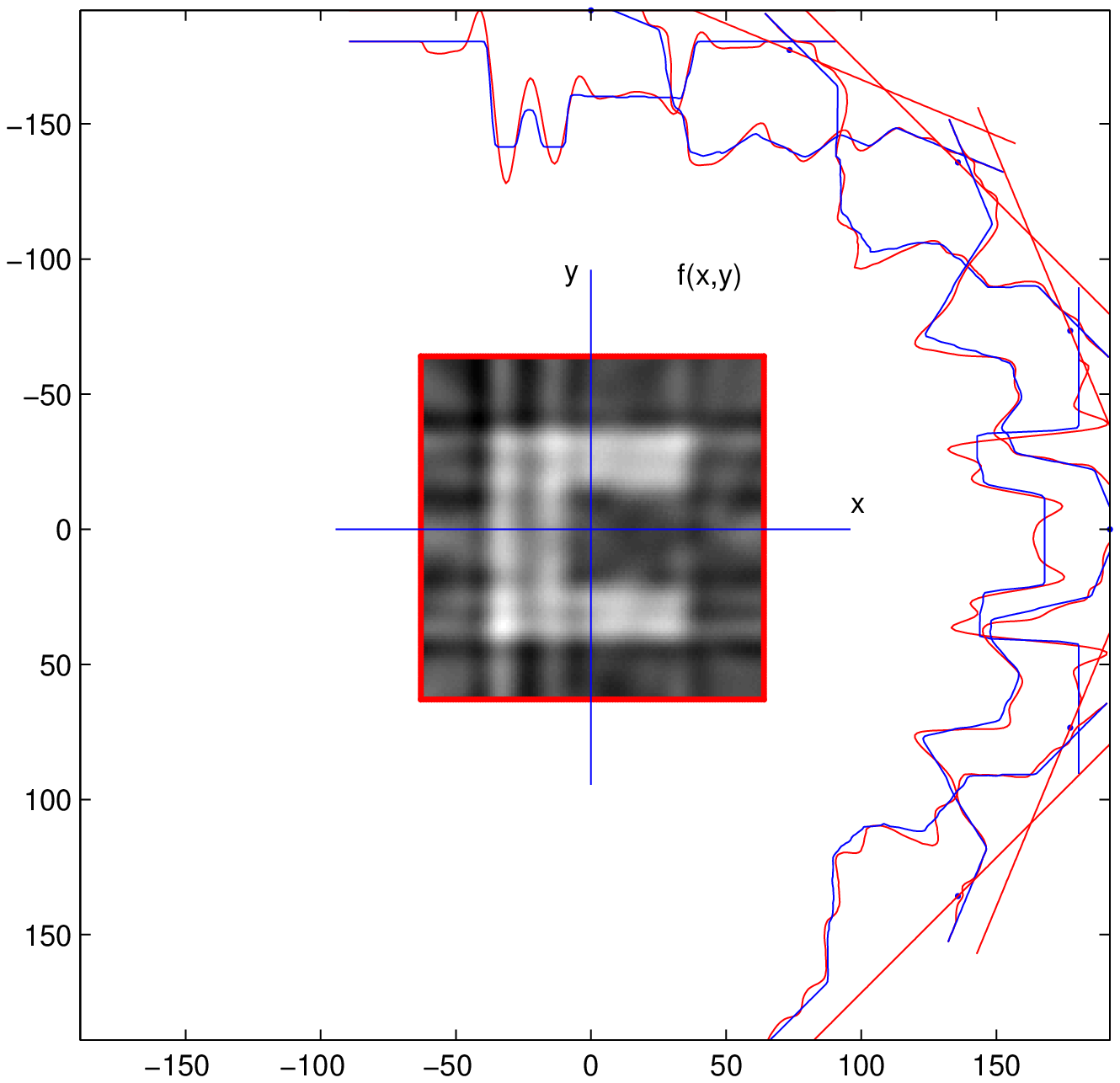}&
d) \includegraphics[width=35mm,height=35mm]{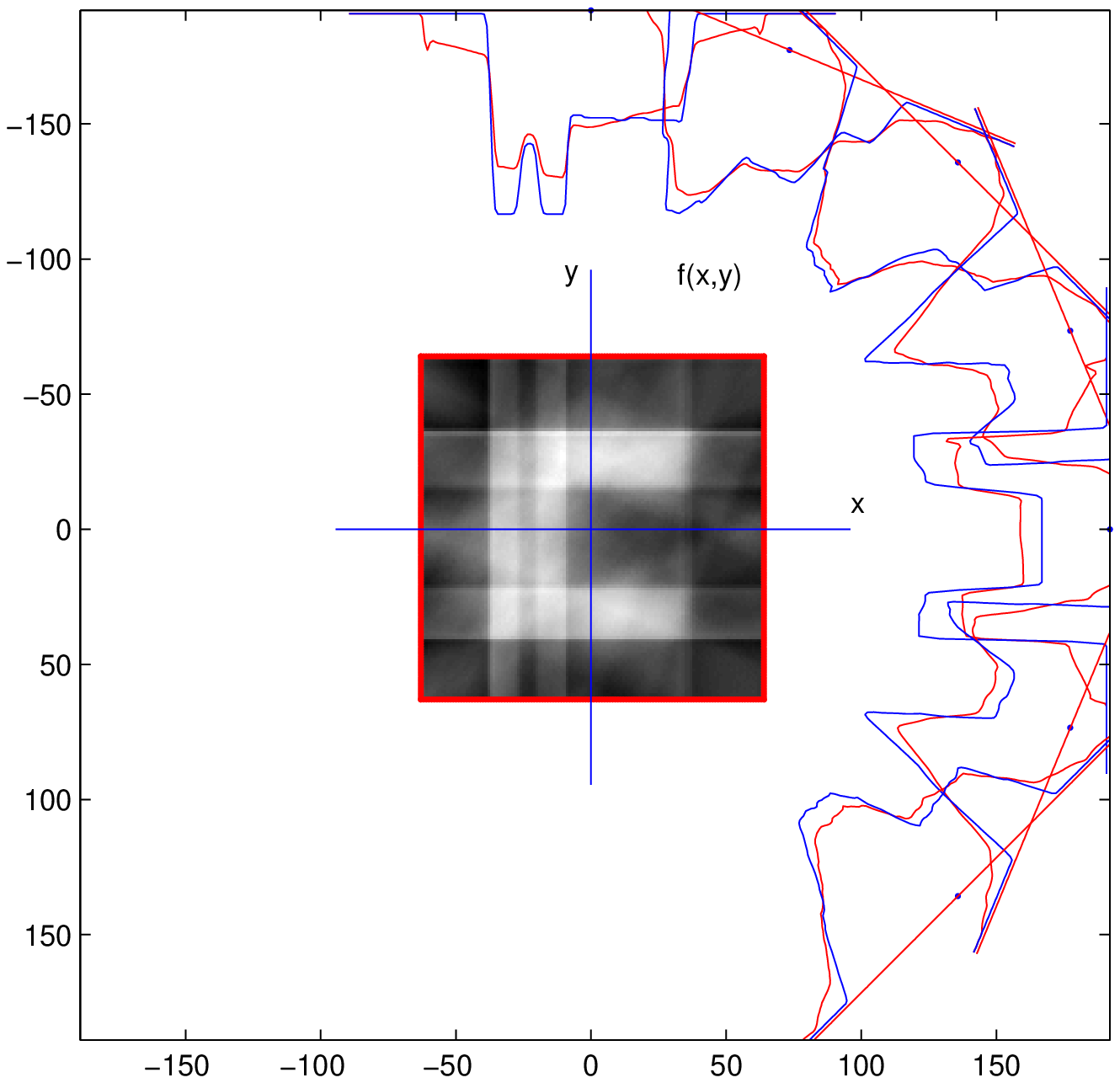}\\[6pt]
e) \includegraphics[width=35mm,height=35mm]{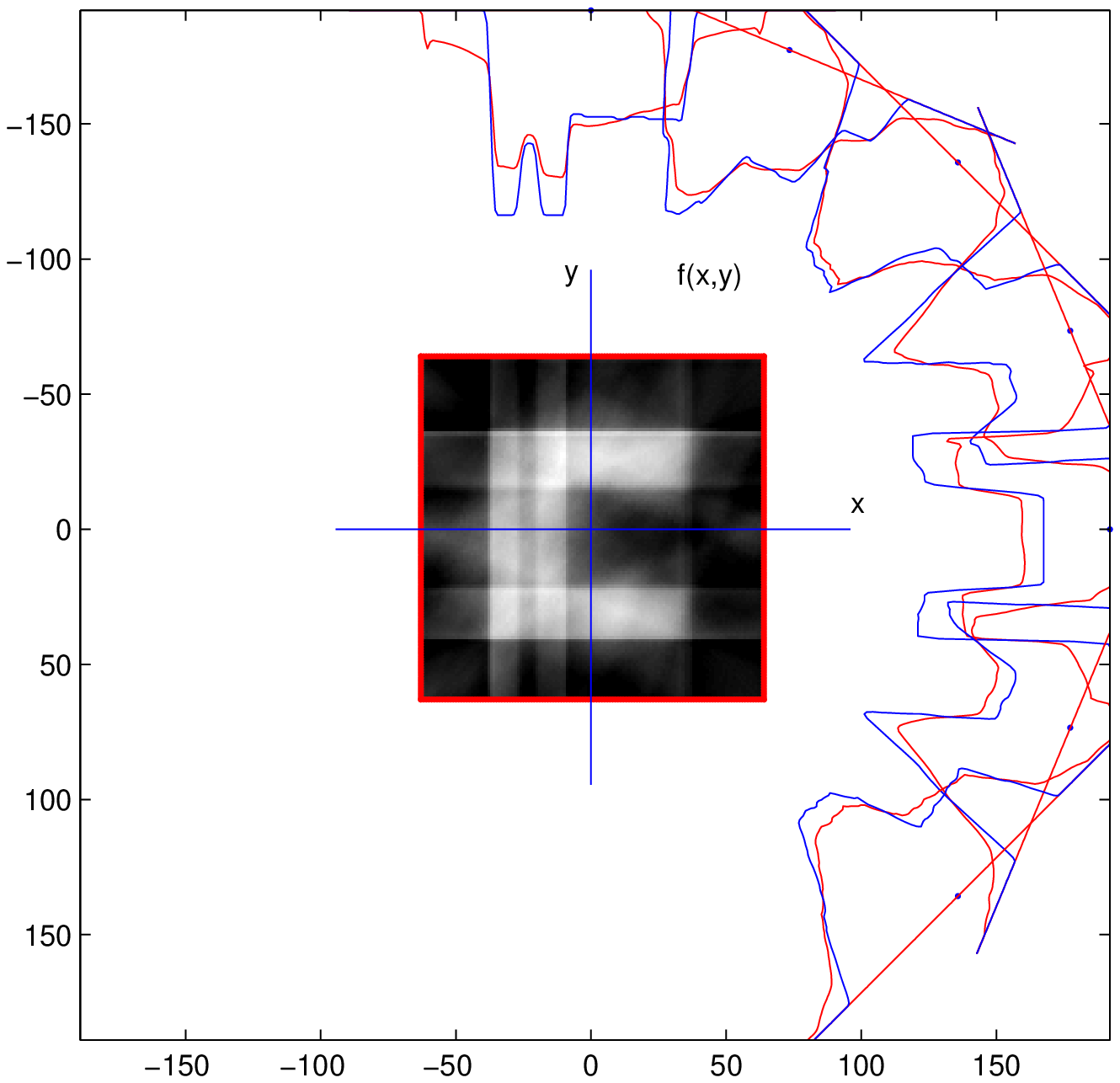}& 
f) \includegraphics[width=35mm,height=35mm]{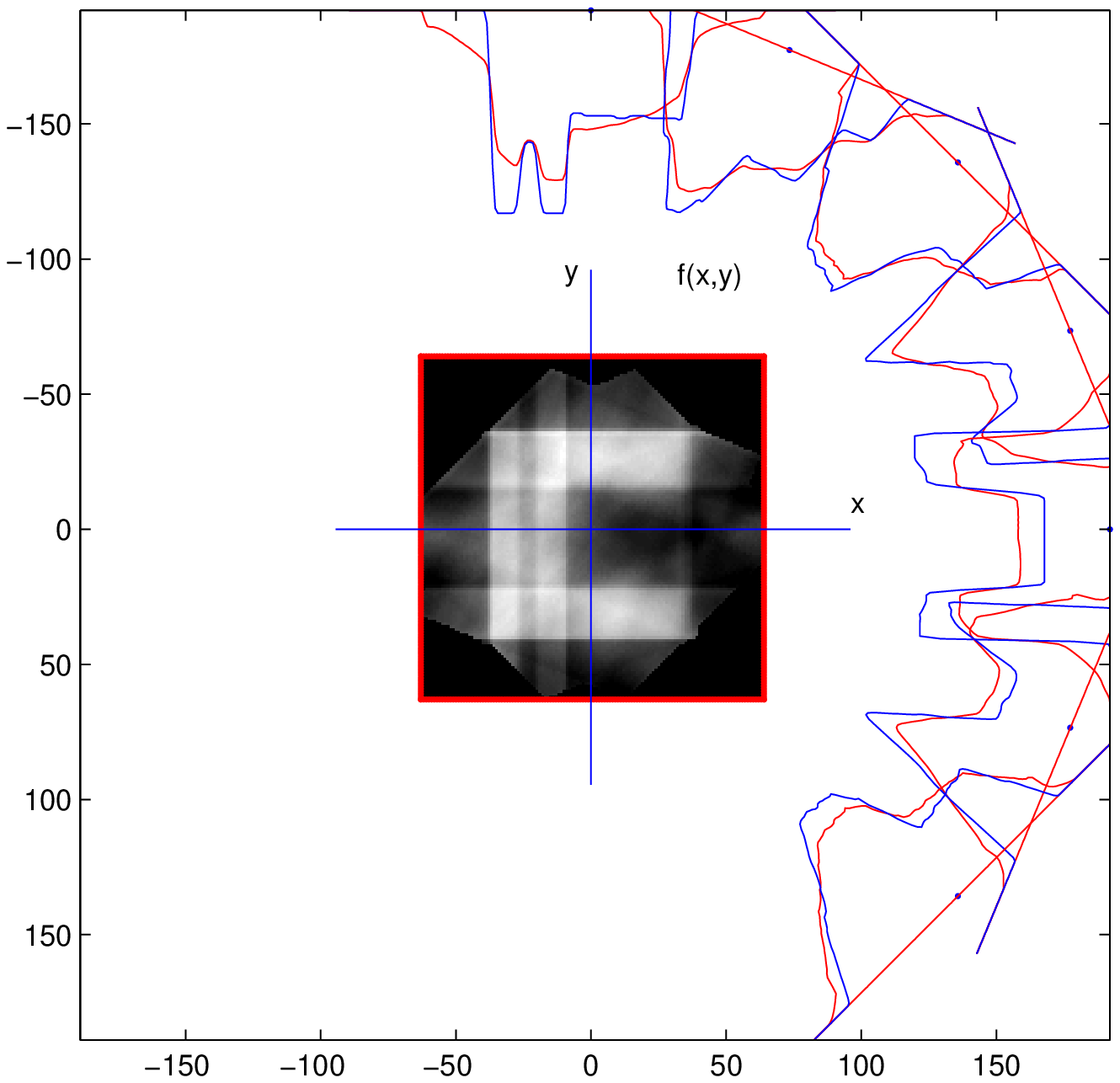}\\[6pt]
g) \includegraphics[width=35mm,height=35mm]{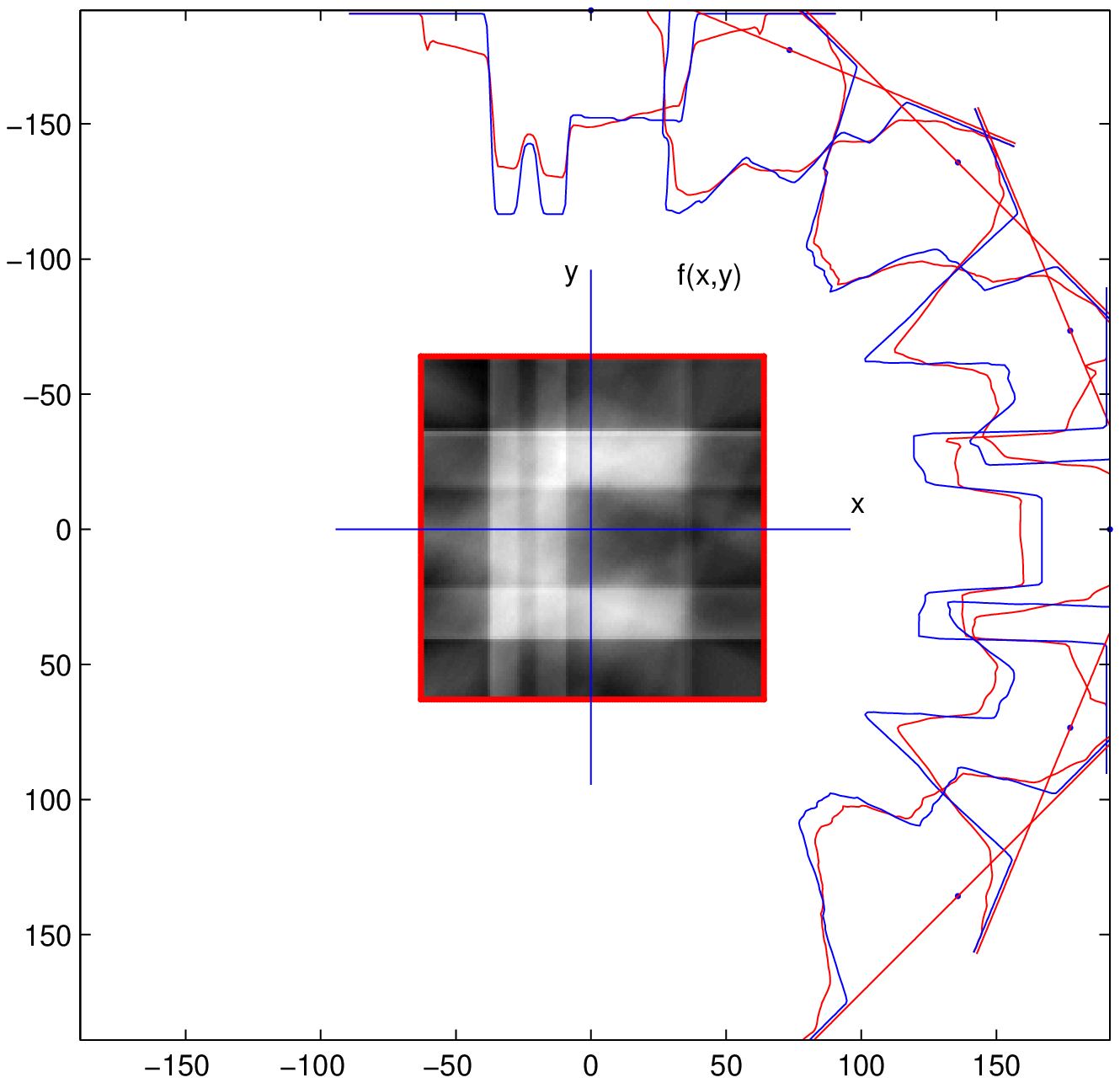}&
h) \includegraphics[width=35mm,height=35mm]{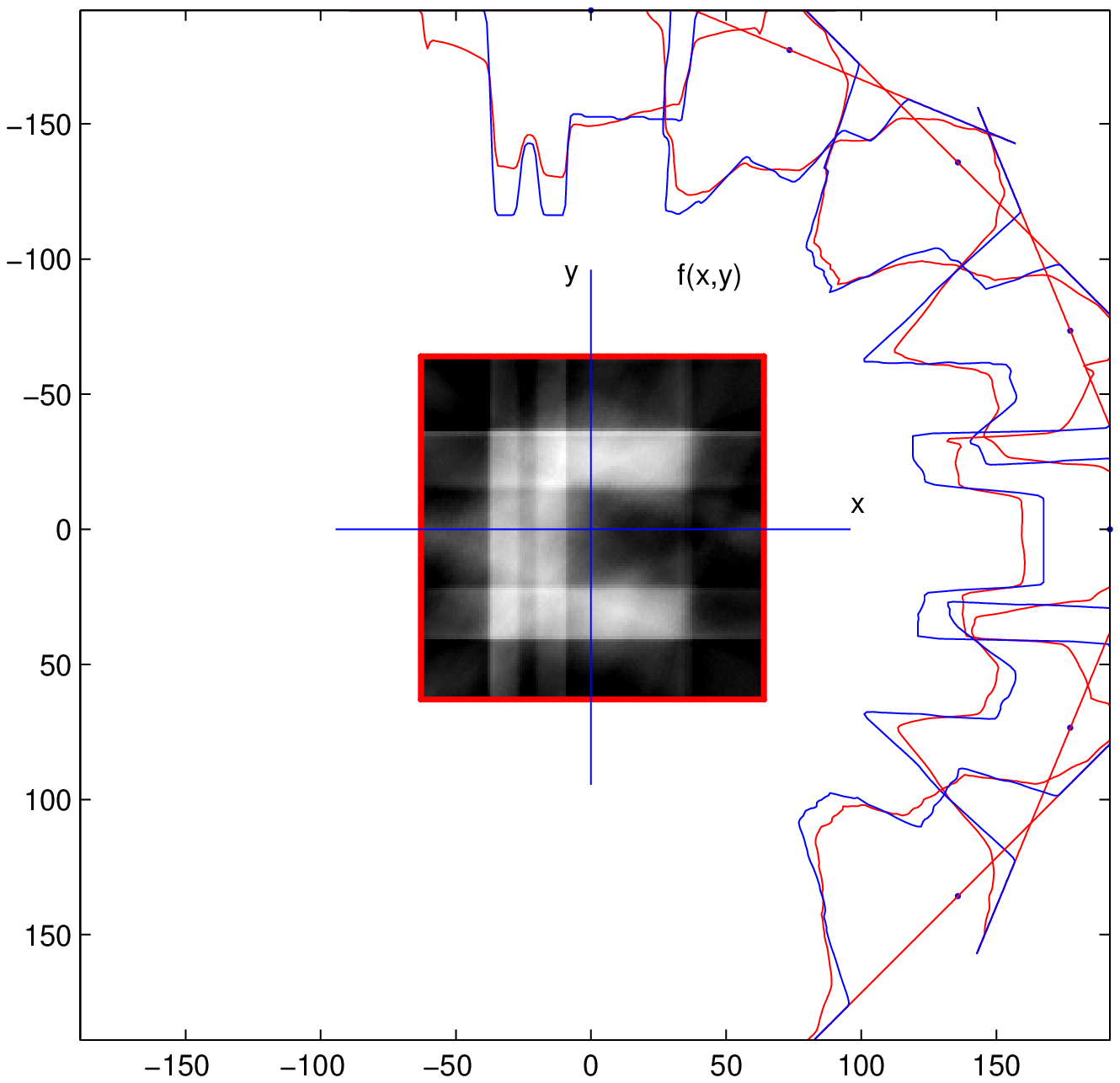}
\etabu
\\
{\bf Fig.~7~:} Reconstruction \`a partir de 7 projections:   
a) l'objet original et les donn\'ees,  b) r\'etroprojection   
c) r\'etroprojection filtr\'e avec contrainte  
d) MC, e) MC avec positivit\'e, 
f) MC avec positivit\'e et contraint de support,  
g) R\'egularisation quadratique (RQ), 
h) RQ avec positivit\'e,
\ecc
}}}

\section{Mod\'elisation par champs de Markov composites}
\'Evidament, plus on a des donn\'ees bien r\'eparties, 
mieux sera les r\'esultats. 
Mais, lorsque l'obtention d'autres projections est impossible, il faudra r\'ecompenser la manque d'information par des mod\'elisations plus pr\'ecises. 
En particulier, dans le domaine du contr\^ol non destructif (CND), une information \aprio importante est que l'objet est compos\'e d'un nombre fini de mat\'eriaux. Ceci signifie que l'image que nous cherchons \`a reconstruire est compos\'ee d'un nombre fini de r\'egions homog\`emes. C'est exactement la mod\'elisation de cette information \aprio qui est l'originalit\'e des travaux que nous menons dans notre laboratoire. 

L'outil est la mod\'elisation probabiliste par champs de Markov et l'estimation bay\'esienne. Un grand nombre de travaux ont \'et\'e fait sur ce sujet (voir par exemple \cite{Nikolova94,Saquib98,Djafari02b}). 
Ici, nous mentionnons seulement deux mod\'elisations~: 
Mod\'elisation de l'image par un champs composite (intensit\'es-contours) ou (intensit\'es-r\'egions). 
Dans la premi\`ere, on introduit une variable cach\'ee binaire $q(\rb)$ qui repr\'esent les contours et dans la deuxi\`eme on introduit une variable cach\'ee discr\`ete $z(\rb)$ qui peut prendre 
des valeurs discr\`etes $k=1,\cdots,K$, repr\'esentant les labels attribu\'es 
aux pixels $f(\rb)$ de l'image ayant les m\^emes propri\'et\'es (par exemple se trouvant dans une m\^eme r\'egion homog\`ene). 

\subsection{Mod\`ele Intensit\'es-Contours}
Dans cette mod\'elisation, l'id\'ee de base est de mod\'eliser le fait qu'une image est en faite une fonction $f(\rb)$ qui est continue par morceaux (piecewise continuous) ou par r\'egions. Il y a donc des discontinuit\'es (contours). On peut alors mod\'eliser ces contours par une image binaire $q(\rb)$. Le point essentiel est alors de d\'ecrire \`a l'aide d'une loi de probabilit\'e conditionnelle $p(\fb|\qb)$, le lien qu'il y entre des variables 
intensit\'es $\fb$ et des variables contours $\qb$ qui peut \^etre r\'esum\'e par~: 
\\ ~\\ 
Cas 1D: 
\[
p(f_j|q_j,f_i, i\not=j)=\Nc(\beta (1-q_j)f_{j-1},\sigma_f^2)
\]
Cas 2D: 
\[ 
p(f(\rb)|q(\rb),f(\sb))=\Nc\left(\beta (1-q(\rb))\sum_{\sb\in\Vc(\rb)} f(\sb), \sigma_f^2\right)
\]
Ensuite, en choisissant une loi \aprio appropri\'ee pour $p(\qb)$ et en utilisant des lois $p(\gb|\fb)$ et $p(\fb|\qb)$, on obtien la loi \apost $p(\fb,\qb|\gb)$ 
qui peut \^etre utilis\'ee pour inferer conjointement sur $\fb$ et sur $\qb$. 
A titre d'indication, consid\'erons l'estimation au sense du MAP~:
\[
(\fbh,\qbh)=\argmax{\fb,\qb}{p(\fb,\qb|\gb)}
\]
qui peut \^etre obtenu par un algorithme it\'eratif du type~:
\[
\barr{l}
\fbh=\argmax{\fb}{p(\fb|\gb,\qb)}=\argmin{\fb}{J(\fb)}
\\
\qbh=\argmax{\qb}{p(\qb|\gb)}
\earr
\]
avec
\[
J(\fb)=\|\gb-\Hb\fb\|^2+\sum_{\rb} (1-q(\rb)) \left(f(\rb)-\beta \sum_{\sb\in\Vc(\rb)} f(\sb)\right)^2
\]
L'\'etape difficile est l'obtention de l'expression 
de $p(\qb|\gb)$ et surtout son optimisation, qui id\'ealement ne peux se faire qu'\`a l'aide d'une recherche combinatoire. Il existe un tr\`es grand nombres de travaux portant sur diff\'erentes approximations qui permettent d'effectuer cette optimisation d'une mani\`ere approch\'ee mais r\'ealiste en co\^ut de calcul pour des applications r\'eelles.  

Pour plus de d\'etail sur cette m\'ethode se r\'ef\'erer \`a \cite{Nikolova94,Saquib98,Djafari02b}. 

Ici, nous montrons un r\'esultat typique que l'on peut obtenir avec de telles m\'ethode.  
Comme nous pouvons constat\'e, cette mod\'elisation \aprio n'est pas encore suffisament forte pour obtenir un r\'esultat satisfaisant pour ce Inverse problem tr\`es difficile. 

\noindent\fbox{\noindent\hbox{\vbox{
\bcc
\btabu{@{}l@{}l@{}}
a) \includegraphics[width=40mm,height=40mm]{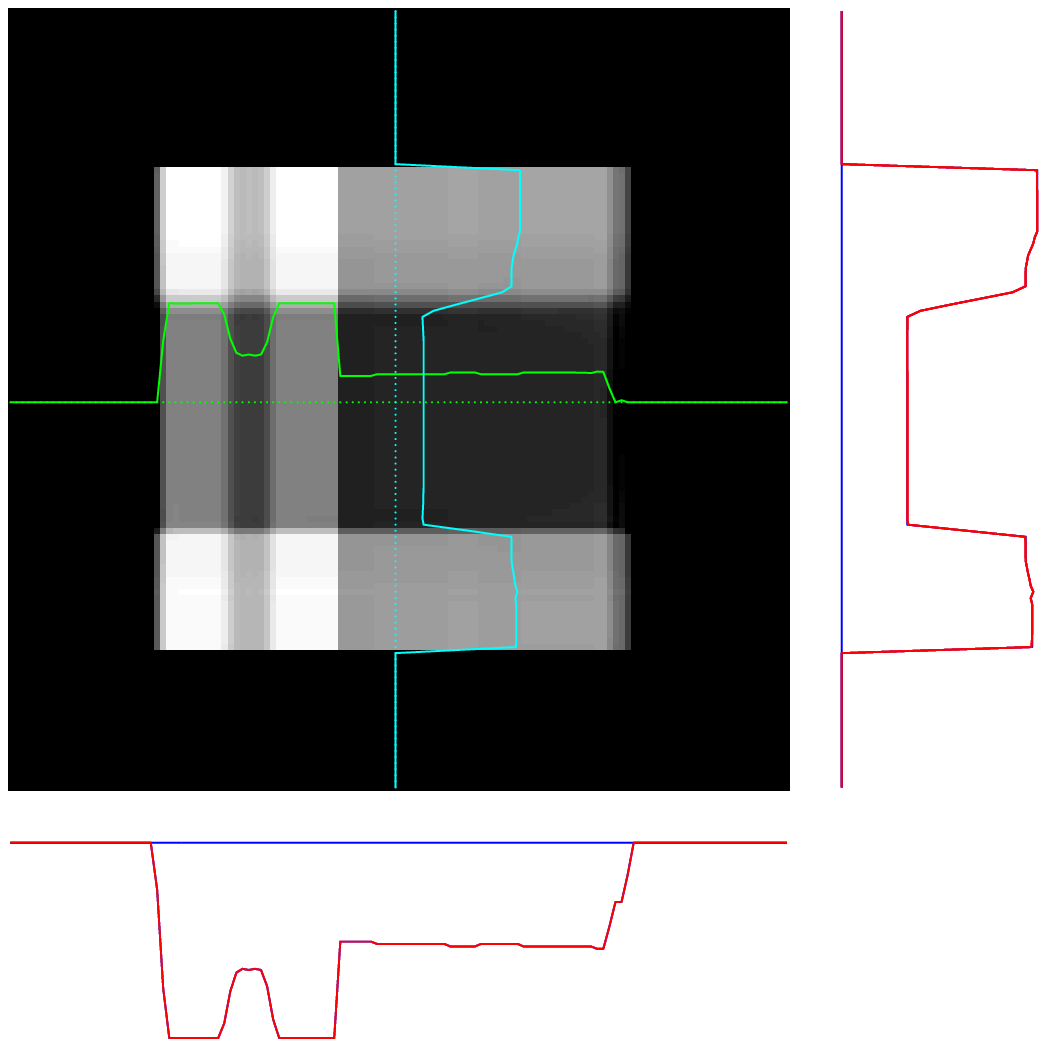}&
b) \btabu[b]{@{}l@{}}
\includegraphics[width=30mm,height=30mm]{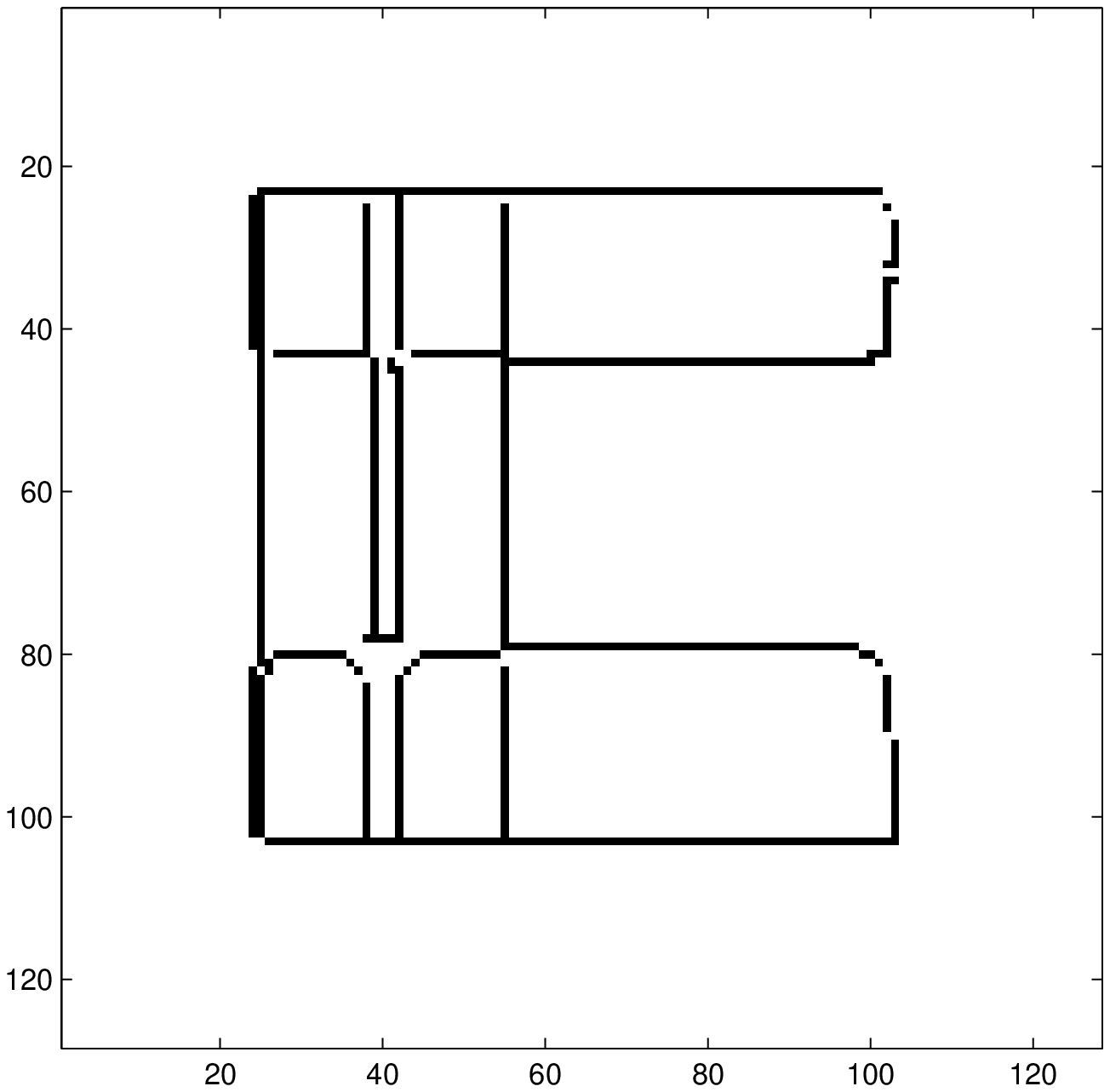}\\[15pt] ~\\ 
\etabu
\etabu
\\
{\bf Fig.~8~:} R\'esultats de l'estimation des intensit\'es $\fb$ en (a) et des contours $\qb$ en (b). 
\ecc
}}}

\subsection{Mod\`ele Intensit\'es-R\'egions}
La mod\'elisation pr\'ec\'edente, bien que d\'ej\`a plus sp\'ecifique, n'apportait pas d'information sur des valeurs qui peuvent prendre des pixels de l'image. 
Dans certaines applications, par exemple en contr\^ol non destructif (CND), 
nous savons \aprio que l'objet est compos\'e d'un nombre fini de mat\'eriaux 
(air, m\'etal, composite). La mod\'elisation qui suit permet de prendre en compte ce type d'information \aprio. 

Plus sp\'ecifiquement, nous proposons de mod\'eliser l'image par un champs composite (intensit\'es-labels), o\`u les labels $z(\rb)$ repr\'esentent la nature de mat\'eriaux \`a la position du pixel $\rb$ et l'homog\'en\'eit\'e des intensit\'es $\fb_k=\{f(\rb), \rb\in\Rc_k\}$ dans une r\'egion donn\'ees $\Rc_k=\{\rb : z(\rb)=k\}$ est mod\'elis\'e par un champs de Gauss-Markov:
\[
p(\fb_k)=\Nc(m_k\oneb,\Sigmab_k)
\]
ce qui peut \^etre interpr\'et\'e aussi par des relations suivantes: 
\[
\barr{l}
p(f(\rb)|z(\rb)=k)=\Nc(m_k, v_k) \\ 
p(f(\rb))=\sum_{k=1}^K p(z(\rb)=k) \; \Nc(m_k,\sigma_k^2) \\ 
\Sigmab_k=\diag{v_1,\cdots,v_K} 
\earr
\]
ce qui montre que la distribution marginale de chaque pixel de l'images 
est mod\'elis\'ee par un m\'elange de gaussiennes. 

Supposant ensuite qu'\aprio les pixels qui se trouvent dans deux r\'egions diff\'erentes soient ind\'ependantes, on peut \'ecrire~:
\[
p(\fb|\zb)=\prod_{k=1}^K \Nc(m_k\oneb,\Sigmab_k)
\]

La particularit\'e de la m\'ethode que nous proposons est une mod\'elization sp\'ecifique pour des labels des r\'egions (Champs de Potts) qui permet 
de d\'ecrire pour $p(\zb)$
\[
p(\zb)\propto
\expf{ \alpha \sum_{\rb\in\Rc} \sum_{\sb\in\Vc(\rb)} \delta (z(\rb)-z(\sb))}
\]

Notant aussi par $\thetab=\{\sigmae^2,(m_k,\sigma_k^2),k=1,\cdots,K\}$ 
que l'on appelle le vecteur des hyperparam\`etres, on aura \`a exprimer la loi 
\apost jointe $p(\fb,\zb,\thetab|\gb)$, qui peut ensuite \^etre utilis\'ee pour estimer ces inconnues. 
Diff\'erents choix sont alors possibles. Ici, nous mentionnons ceux que nous avons impl\'ement\'es et utilis\'es:
\bit
\item MAP (Algorithm 1):
\[
\left\{\barr{rlll}
\fbh 
&=\argmax{\fb}{p(\fb|\zb,\thetab,\gb)}\\
\thetabh 
&=\argmax{\thetab}{p(\thetab|\fb,\zb,\gb)} 
%&\mbox{ou~} =\argmax{\thetab}{p(\thetab|\zb,\gb)}
\\ 
\zbh 
&=\argmax{\zb}{p(\zb|\fb,\thetab,\gb)} 
%&\mbox{ou~} =\argmax{\zb}{p(\zb|\thetab,\gb)} 
\earr\right.
\]

\item MAP-Gibbs (Algorithm 2):
\[
\left\{\barr{rlll}
\fbh 
&=\argmax{\fb}{p(\fb|\zb,\thetab,\gb)} \\ 
\mbox{\'echant.~~} \thetabh 
&\mbox{avec~~} {p(\thetab|\fb,\zb,\gb)} 
%&\mbox{or with~~} {p(\thetab|\zb,\gb)} 
\\
\mbox{\'echant.~~} \zbh 
&\mbox{avec~~} {p(\zb|\fb,\thetab,\gb)} 
%&\mbox{or with~~} {p(\zb|\thetab,\gb)}
\earr\right.
\]
\eit
Pour plus de d\'etails sur les expressions de ces lois et la mise en oeuvre de la m\'ethode dans un cadre plus g\'en\'eral se r\'ef\'erer \`a 
\cite{snoussi04b,snoussi04c,feron05a,humblot05a}. 

Principal avantage d'une telle mod\'elisation et d'un tel m\'ethode est que 
l'on obtient non seulement une estimation de $\fb$ mais aussi de $\zb$ qui repr\'esente une segmentation de l'image, et aussi par un simple algorithme de d\'etection de contours sur $\zb$ on obtiendrai aussi une image des contours $\qb$. La figure qui suit montre un r\'esultat typique.

\medskip
\noindent\fbox{\noindent\hbox{\vbox{
\bcc
\btabu{@{}l@{}l@{}}
\includegraphics[width=40mm,height=40mm]{f00}&
\includegraphics[width=40mm,height=40mm]{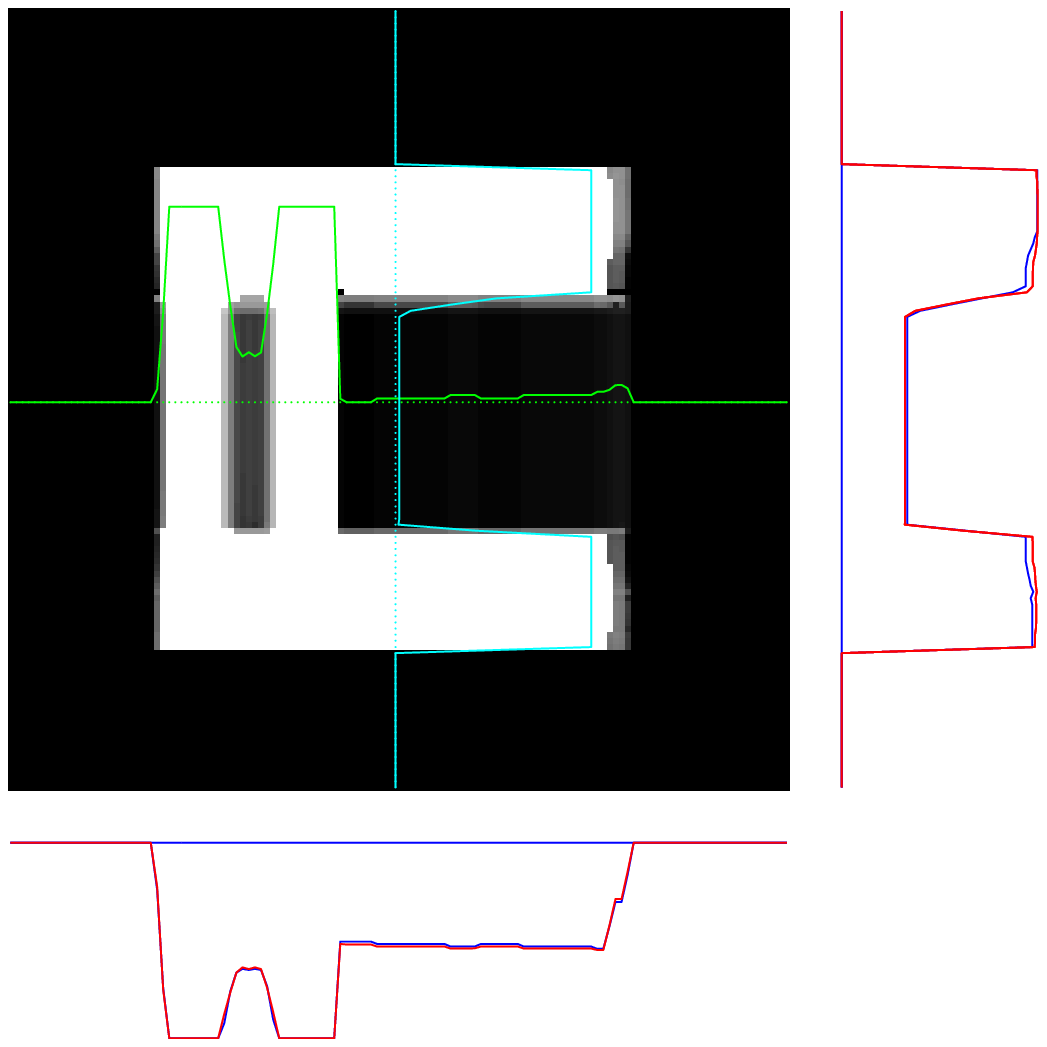}\\
a & b \\
\includegraphics[width=30mm,height=30mm]{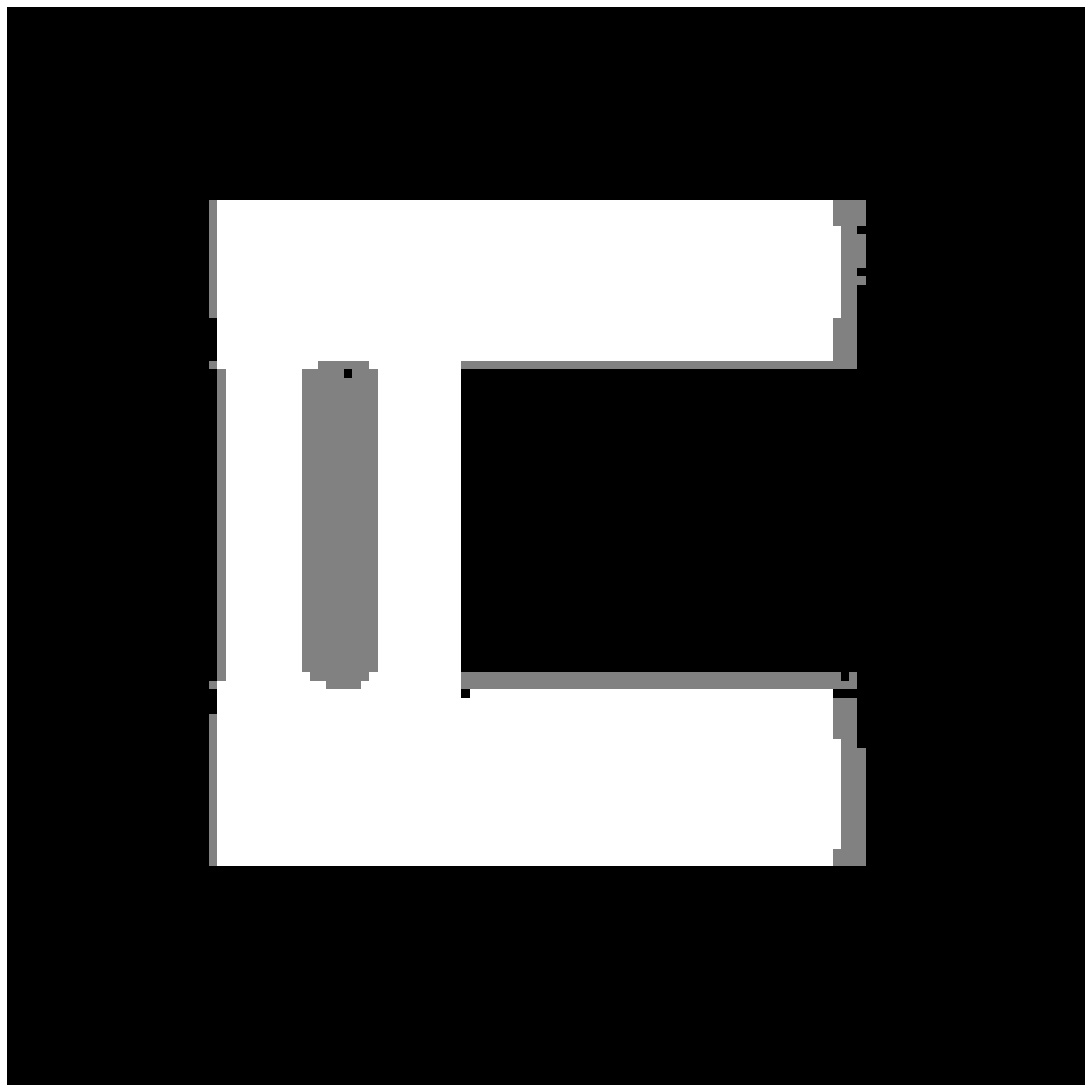}&
\includegraphics[width=30mm,height=30mm]{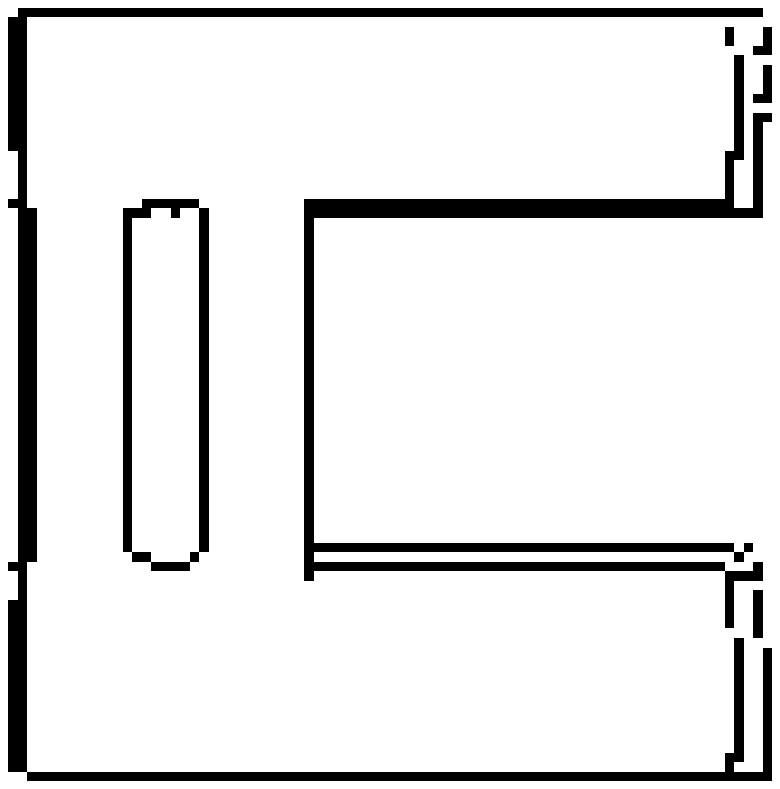}\\[3pt]  
c & d 
\etabu
\\
{\bf Fig.~8~:} a) Objet original et les deux projections, 
b,c et d) R\'esultats de l'inversion par la m\'ethode propos\'ee qui fourni 
une estimation des intensit\'es (b), une estimation de labels en (c) et une estimation des contours en d). 
\ecc
}}}

%%%%%%%%%%%%%%%%%%%%%%%%%%%%%%%%%%%%%%%%%%%%%%%%%%%%%%%%%%%%%%%%%%
\section{Conclusion}
%%%%%%%%%%%%%%%%%%%%%%%%%%%%%%%%%%%%%%%%%%%%%%%%%%%%%%%%%%%%%%%%%%
Dans ce travail, \`a but p\'edagogique, au travers d'un 
Inverse problem de la reconstruction d'image en Tomography X lorsque le nombre de projections sont tr\`es limit\'e, nous avons analys\'e les difficult\'es inh\'erentes des probl\`emes inverses. Le principal objectif \'etait de montrer que les diff\'erentes m\'ethodes classiques na\^ives, mais tr\`es utilis\'ees, ne donnent pas de solutions satisfaisantes et qu'il y a un besoin de proposer des m\'ethodes d'inversion plus sophistiqu\'ees qui permettent d'introduire de l'information \aprio n\'ecessaire pour compenser la manque d'information dans les donn\'ees. 

Un grand nombre de mod\'elisations ont \'et\'e propos\'ees (voir par exemple \cite{Kuba99,Soussen04a}). Mais, ici, nous nous sommes content\'e des m\'ethodes 
qui mod\'elisent l'image au niveau des pixels. 

Le cas particulier de la reconstruction \`a partir de deux projections est 
d\'etaill\'e et une m\'ethode bas\'ee sur la mod\'elisation de l'image par un champs de Markov compos\'e (intensit\'e-labels) et l'estimation bay\'esienne est pr\'esent\'ee qui permet, au moins, d'obtenir une solution satisfaisante au probl\`eme.  
Les routines Matlab correspondant est disponible sur {\tt http://djafari.free.fr/TomoX}.
%%%%%%%%%%%%%%%%%%%%%%%%%%%%%%%%%%%%%%%%%%%%%%%%%%%%%%%%%%%%%%%%%%

{\small 
\def\sca#1{{\sc #1}}
\bibliographystyle{ieeetr}
\bibliography{bibenabr,revuedef,revueabr,baseAJ,baseKZ,gpipubli,/home/djafari/Tex/Inputs/bib/amd/amd_art,/home/djafari/Tex/Inputs/bib/amd/amd_ch}
}

\edoc